\documentclass{aa}  
\usepackage{orcidlink}
\usepackage{graphicx}
\usepackage{txfonts}
\usepackage[normalem]{ulem}
\usepackage{hyperref}

\usepackage{color}
\definecolor{amethyst}{rgb}{0.8, 0.0, 0.0}

\definecolor{orange}{rgb}{1,0.5,0}

\definecolor{boh}{rgb}{1,0,0}

\definecolor{mile}{rgb}{0.5,0,0.5}

\newcommand{\rev}[1]{#1}
\newcommand{\revv}[1]{{#1}}

\def\mtwo{M_{\rm 200}}
\def\rtwo{R_{\rm 200}}
\def\msun{\rm{M_\odot}}

\newcommand{\hmol}{\,H$_{2}$\;}

\begin{document}

   \title{Intertwined formation of $\rm{H_2}$, dust, and stars in cosmological simulations}

   \subtitle{}

   \author{Cinthia Ragone-Figueroa \inst{1,2,3}   \orcidlink{0000-0003-2826-4799}
          \and Gian Luigi Granato   \inst{2,1,3} \orcidlink{0000-0002-4480-6909}
          \and Massimiliano Parente     \inst{4,2,3}   \orcidlink{0000-0002-9729-3721}
          \and
          Giuseppe Murante \inst{2,3,7}
          \and
          Milena Valentini \inst{5,2,6,7}  \orcidlink{0000-0002-0796-8132}
          \and
          Stefano Borgani\inst{5,2,3,6,7}
          \and
          Umberto Maio
          \inst{2,3}
          }

   \institute
   {
   IATE - Instituto de Astronom\'ia Te\'orica y Experimental, Consejo Nacional de Investigaciones Cient\'ificas y T\'ecnicas de la\\ Rep\'ublica Argentina (CONICET), Universidad Nacional de C\'ordoba, Laprida 854, X5000BGR, C\'ordoba, Argentina\\
   \email{cinthia.ragone@unc.edu.ar}
   \and   
    INAF, Osservatorio Astronomico di Trieste, via Tiepolo 11, I-34131, Trieste, Italy
    \and
    IFPU, Institute for Fundamental Physics of the Universe, Via Beirut 2, 34014 Trieste, Italy
   \and
   SISSA, Via Bonomea 265, I-34136 Trieste, Italy     
   \and 
    Dipartimento di Fisica dell'Universit\`a di Trieste, Sez. di Astronomia, via Tiepolo 11, I-34131 Trieste, Italy 
   \and
   INFN, Instituto Nazionale di Fisica Nucleare, Via Valerio 2, I-34127, Trieste, Italy
   \and
   ICSC - Italian Research Center on High Performance Computing, Big Data and Quantum Computing, via Magnanelli 2, 40033, Casalecchio di Reno, Italy
   }

   \date{Received September 15, 2024; accepted March 16, 2024}

 
  \abstract
   {Molecular hydrogen ($\rm{H_2}$) plays a crucial role in the formation and evolution of galaxies, serving as the primary fuel reservoir for star formation. In a metal-enriched Universe, $\rm{H_2}$ forms mostly through catalysis on interstellar dust grain surfaces. However, due to the complexities of modelling this process, star formation in cosmological simulations often relies on empirical or theoretical frameworks that have only been validated in the local Universe to estimate the abundance of $\rm{H_2}$.  
   }
   {The goal of this work is to model the connection between the processes of star, dust, and $\rm{H_2}$ formation in our cosmological simulations.}
   {Building upon our recent integration of a dust evolution model into the star formation and feedback model MUPPI, we included the formation of molecular hydrogen on the surfaces of dust grains. We also accounted for the destruction of molecules and their shielding from harmful radiation. 
   }
   {The model reproduces, reasonably well, the main statistical properties of the observed galaxy population for the stellar, dust, and \hmol components. The evolution of the molecular hydrogen cosmic density ($\rho_{\rm{H2}}$) in our simulated boxes peaks around redshift $z=1.5$, consistent with observations. Following its peak, $\rho_{\rm{H2}}$ decreases by a factor of two towards $z=0$, which is a milder evolution than observed. Similarly, the evolution of the molecular hydrogen mass function since $z=2$ displays a gentler evolution when compared to observations. Our model recovers satisfactorily the integrated molecular Kennicut-Schmidt (mKS) law between the surface star formation rate ($\Sigma_{\rm SFR}$) and surface \hmol density ($\Sigma_{\rm H2}$) at $z=0$. This relationship is already evident at $z=2$, albeit with a higher normalization. We find hints of a broken power law with a steeper slope at higher $\Sigma_{\rm H2}$. We also study the $\rm{H_2}$-to-dust mass ratio in galaxies as a function of their gas metallicity and stellar mass, observing a decreasing trend with respect to both quantities. The $\rm{H_2}$-to-dust mass fraction for the global population of galaxies is higher at higher redshift. The analysis of the atomic-to-molecular transition on a particle-by-particle basis suggests that gas metallicity cannot reliably substitute the dust-to-gas ratio in models attempting to simulate dust-promoted $\rm{H_2}$.
   }
   {}


   \maketitle
%

\section{Introduction}
Despite the substantial progress achieved in the last decades, both observationally and theoretically driven, 
a complete understanding of the various processes involved in galaxy formation has yet to be reached. Among them, modelling the pivotal phenomenon of star formation (SF) faces formidable challenges arising from the intricate interplay of complex physical mechanisms acting across a wide range of scales, from the large-scale structure (of the order of megaparsec) to the sub-stellar ones ($\lesssim 10^{-8}$~pc). The ultimate reservoir for SF is the densest and coldest phase of the interstellar medium (ISM), dominated in mass by molecular hydrogen (${\rm H_2}$) and organized in self-gravitating dusty molecular clouds (MC)
featuring an approximate mass range from about $10^4$ to $10^6$~$\msun$
\cite[e.g.][]{KennicuttEvans2012,miville17}. 
The gravitational collapse of these clouds initiates the cooling and SF processes within them. Indeed, while the observed star formation rate (SFR) surface density of galaxies correlates with the neutral gas surface density \citep[the Schimdt-Kennicutt law e.g.][]{schmidt59, Kennicutt1998}, the correlation becomes, unsurprisingly, much stronger when only the molecular gas surface density is considered \citep[e.g.][]{wong_blitz_02, Bigiel2008}. 
While the most widespread interpretation of the latter correlation is that the formation of molecular gas controls the SFR, it has also been proposed that both of these processes, the MC formation and the onset of SF, might share a common underlying cause \citep[e.g.][]{krumholz2011,maclow12}.

Numerical simulations of galaxy formation in a cosmological context are far from the resolution required to resolve the structure of MC, the minimal requirement to model SF from first principles. Moreover, they do not incorporate many of the physical mechanisms that determine the MC evolution to form stars. Instead, they rely on various proposed sub-resolution prescriptions that connect the SFR with the estimated availability of gas eligible for SF. In most cases, this simply means considering gas that, at a resolved level, turns out to be cold and dense enough ($T \lesssim \mbox{a few} \times 10^4$~K; $n\gtrsim 0.1$-$10$ cm$^{-3}$), and that the local SFR is assumed to be proportional to its density divided by an SF timescale usually proportional to the local \rev{free-fall time}. The Schimdt-Kennicutt law loosely inspires the latter relationship.
%
%
This kind of approach is usually adopted when simulating large cosmological volumes. Examples include Illustris \citep[][]{Vogelsberger2014}, Magneticum \citep[][]{Dolag2015}, NIHAO \citep[][]{Wang2015}, IllustrisTNG \citep[][]{Pillepich2018}, the Massive-Black \citep[][]{Khandai2015} and Fable \citep[][]{Henden2018} simulations, the Horizon-AGN suite \citep[][]{Dubois2016, Dubois2021} and the DIANOGA simulations \citep[][]{ragone18,Bassini2020}. 

Somewhat more elaborated treatments compute the local SFR by including a sub-resolution prescription to estimate the fraction of cold and dense gas that is in molecular form. The prescription can be derived by theoretical modelling \citep[e.g.][]{Krumholz2008, Krumholz2009, KrumholzGnedin2011}, as in \cite{Kuhlen2012},  the Mufasa and Simba simulations \citep[][]{Dave2017, Dave2019}, and the FIRE suite \citep[][]{Hopkins2014, Hopkins2018}, or observations \citep[e.g.][]{Blitz2006}, as in \cite{murante15}. \footnote{\rev{The same prescriptions have also been used to estimate  in post-processing the \hmol predicted by  simulations, in which, however, the information is not exploited during the run \citep[e.g.][]{lagos2015,popping2019,manuwal2023}}}

At a higher level of complexity, some simulations, thus far dealing with smaller volumes and/or single halos, predict the amount of molecular gas following the main processes of \hmol formation and destruction and take advantage of this prediction to compute the SFR \citep[e.g.][]{Pelupessy2006, Gnedin2009, Christensen2012, Tomassetti2015}. The dominant \hmol formation mechanisms to consider in metal-enriched environments are catalytic reactions occurring on the surfaces of dust grains \citep[for a review see][]{wakelam17}. Direct formation in the gas phase does indeed turn out to be very inefficient due to the difficulty of the axis-symmetric-forming molecule in radiating away the excess energy to remain stable. This process becomes relevant only when the ISM is extremely dust-poor. 
\rev{\citep[e.g.][]{gould_salpeter_63, Krumholz2009, Bekki2013, bron14, Tomassetti2015, valdivia16, maio22}}. 
However, the aforementioned simulation works lack comprehensive dust creation and destruction modelling. Instead, they rely on a fixed dust-to-gas (D/G) ratio linearly scaling with metallicity to estimate the dust content roughly. 
A further complication arises from the dust-catalyzed \hmol formation rate being proportional to the grain's surface rather than its mass. Therefore, it would be relevant to have information on the dust size distribution to estimate it.
The primary \hmol destruction mechanism is believed to be photo-dissociation via the two-step Solomon process caused by photons in the Lyman–Werner (LW) band $912 \textup{~\AA} < \lambda < 1108 \textup{~\AA}$. It is worth pointing out that dust also substantially contributes to shield regions where \hmol forms efficiently on grains from the destructive effect of LW radiation emitted by OB stars.

Clearly, dust plays a pivotal role in determining the \hmol content of the ISM and, consequently, the reservoir available for SF. However, despite its significance, current cosmological simulations have yet to integrate the simultaneous evolution of dust content and \hmol promoted by dust. Prior studies in this field, including \hmol formation on grain surfaces, have adopted a simplistic method, merely scaling the dust content with gas metallicity while overlooking the impact of grain size distribution on dust surfaces. Notably, \cite{romano22} have conducted a simulation which accounts for both dust content and \hmol concurrently. However, it focuses solely on one isolated disc galaxy and spans a relatively short period of 2 billion years of evolution, during which the SFR is not linked to the molecular content.

In this work, we present an update of the MUPPI (MUlti Phase Particle Integrator) sub-resolution model \citep{murante15}, in which the evolution of dust content, molecular ($\rm{H_2}$) gas and SF are mutually linked. Previous versions of MUPPI featured an advanced description of a multi-phase ISM and proved successful in zoomed-in simulations of late-type galaxies \citep[e.g.][and references therein]{valentini20, granato21, valentini23}. The molecular gas content, and thus the SFR, was estimated adopting the \cite{Blitz2006} or \cite{Krumholz2009} laws. Here, we take instead advantage of the treatment of dust evolution introduced by \cite{granato21} and \cite{parente22} to predict the evolution of \hmol, which in turn is used to estimate the local SFR.\\
In \cite{granato21}, we included in MUPPI a treatment of dust formation and evolution and studied the dust properties of a Milky Way (MW)-like galaxy.  
This model allows us to predict dust grains' chemical composition and size distribution, utilizing the \cite{hirashita15} two-size approximation. Furthermore, we have accounted for hot gas cooling due to collisions of ions with dust particles, following the \cite{dwek81} model.
More recently, in \cite{parente22}, we used MUPPI and its treatment of dust evolution for the first time to simulate a cosmological box and test its performance by comparing the simulated galaxy properties with observational constraints concerning a broader population of galaxies. However, in that work, we deliberately adopted the MUPPI modelling, as previously calibrated, by using zoom-in simulations of a MW-like galaxy without modifications. Substantial tensions regarding low dust content at high redshift and low star formation rate density (SFRD) were reported in that work. The dust evolution modelling, in that case, originated only minor modifications to the behaviour of MUPPI related to its effects on cooling. Conversely, introducing a treatment of molecular gas formation promoted by dust and its use as a reservoir for SF makes the treatment of dust deeply intertwined with the core of MUPPI. Therefore, in the present work, we have revised some aspects of its modelling and parameters, obtaining a simulation set that, at the same time, essentially solves the tensions pointed out by \cite{parente22} in their cosmological volumes (Section \ref{sec:resstars}) while preserving the good results discussed in previous papers on zoom-in simulations of MW-like galaxies. 

The structure of the paper unfolds as follows: In Section \ref{sec:simulations}, we provide a concise overview of the numerical simulations, the sub-resolution model MUPPI, and the adjustments implemented in this iteration of the code to simulate a cosmological box reflecting galaxy properties consistent with observations. Section \ref{sec:MUPPIH2} describes the modelling for the formation of \hmol on dust grain surfaces and the approximation employed to simulate the molecule destruction resulting from LW radiation. Our findings, showcasing a satisfactory replication of observed properties through our model galaxies, including the \hmol mass function and the cosmic evolution of the \hmol mass density, are presented in Section \ref{sec:results1}. In Sections \ref{sec:KS}, \ref{sec:h2dust} and \ref{sec:moltrans}, we delve into the analysis of three applications of our model: the integrated molecular Kennicutt-Schmidt (mKS) relation, the determination of \hmol mass from dust, and the transition from atomic to molecular gas. Finally, our conclusions are presented in Section \ref{sec:conclusions}.

\section{Numerical simulations}
\label{sec:simulations}
\begin{figure}
\centering
\includegraphics[width=\columnwidth]{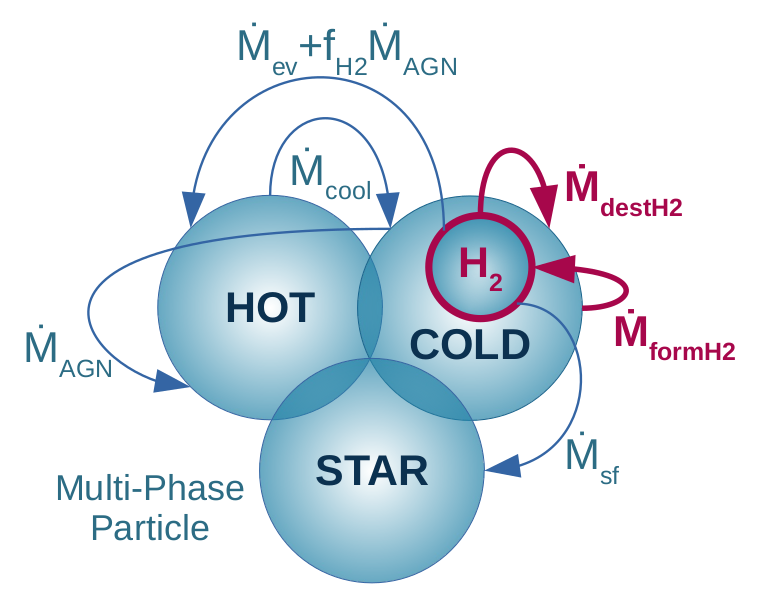}
\caption{Schematic view of the mass fluxes in a multi-phase gas particle. Gas cooling moves mass from the hot to the cold phase at a rate $\dot{M}_{\rm{cool}}$. 
Within the cold phase, an \hmol reservoir begins to grow.
$\dot{M}_{\rm{formH_2}}$ is the \hmol formation rate determined by the particle dust content and the number density of atomic hydrogen, which in turn is computed from the cold phase mass. $\dot{M}_{\rm{destH_2}}$ is the \hmol destruction rate by LW photons, which transports mass from the molecular reservoir back to the cold phase. 
$\dot{M}_{\rm{sf}}$ is the rate at which molecular gas creates a virtual stellar phase, which is contained within the gas particle until a stellar particle is stochastically spawned. Further destruction of the molecular mass occurs due to the action of massive stars and AGN, at rates $\dot{M}_{\rm{ev}}$ and  ${\rm f_{H2}}\ \dot{M}_{\rm{AGN}}$, respectively, where ${\rm f_{H2}}$ represents the fraction of \hmol in the cold phase. AGN also moves material from the cold to the hot phase at a rate $\dot{M}_{\rm{AGN}}$}
\label{fig:MPparticle}
\end{figure}

The sub-resolution model that we employ to account for the unresolved SF, stellar feedback processes and treatment of AGN feedback is MUPPI \citep[MUlti Phase Particle Integrator,][]{murante10,murante15,valentini17,valentini19,valentini20,valentini23}, with some modifications detailed in Section \ref{sec:MUPPImodifications}. 
MUPPI is implemented in the GADGET3 code, a customized version of GADGET2 \citep{springel05}. Our version includes the improved SPH (smoothed particle hydrodynamics) formulation proposed by \cite{beck16}. 
MUPPI assumes that every SPH particle capable of performing SF embodies a multi-phase (MP) ISM, characterized by a hot, a cold, and a virtual stellar phase.
The mass and energy flows among the different components, as depicted in Fig. \ref{fig:MPparticle}, are governed by a set of ordinary differential equations.
A gas particle becomes MP when its number density exceeds 0.005 cm$^{-3}$ and its temperature falls below $5\times10^4$~K. Entering MP does not imply the immediate onset of SF. Instead, it indicates that the gas particle is considered to model the ISM. Once in MP, the hot phase coexists in pressure equilibrium with the cold phase, assumed to maintain a constant temperature of $\rm{T_c=300}$~K.

\subsection{Initial conditions and resolution}
\label{sec:inicond}
To ensure robustness in our results and a sampling of galaxy population sufficient for our purposes, we simulate five cosmological boxes measuring $26\ \text{cMpc}$ per side, from distinct initial conditions\footnote{The initial conditions were generated at $z=99$ using the public code N-GenIC (https://www.h-its.org/2014/11/05/ngenic-code).}. 
\revv{We checked that the z=0 halo mass function is satisfactorily sampled up to $2 \times 10^{13} \msun$ by our total simulated volume.}
The cosmological parameters are taken from \cite{PlanckCollaborationXIII2016}: $\Omega_{\rm m}=\Omega_{\rm DM}+\Omega_{\rm b}=0.3089$, $\Omega_{\rm b}=0.0486$ and $\Omega_\Lambda=0.6911$, $h=0.6774$. 
The power spectrum has a primordial index $n_s=0.9667$ and normalization $\sigma_8=0.8159$.

We evolve $256^3$ DM particles and $256^3$ gas particles. The mass resolution of DM particles is $ \simeq 3.75 \times 10^7~\msun$, while gas particles have an initial mass $ \simeq 7 \times 10^6 ~\msun$. 
For the computation of the gravitational force, we use a Plummer-equivalent softening length of 960 pc, constant in comoving units for $z>6$ and constant in physical units at lower redshift. 

\rev{Our simulations achieve a resolution similar to that employed in the largest boxes of state-of-the-art suites such as Eagle \citep[][]{Schaye2015}, IllustrisTNG \citep{Pillepich2018}, and Simba \citep{Dave2019}, which is relatively modest. In this work, we compare our results with globally integrated properties of galaxies derived from masses in the various components and overall sizes. However, \cite{murante15} showed that the advanced sub-resolution model MUPPI allows to capture to some level the structure and dynamics of MW-size galaxies, even at the resolution of their AqC6 and GA1 initial conditions (readers can refer to their table 1 and figures 17 and 18), which is very similar to that adopted in this work.}

\begin{figure*}
\centering
\includegraphics[width=0.16\textwidth]{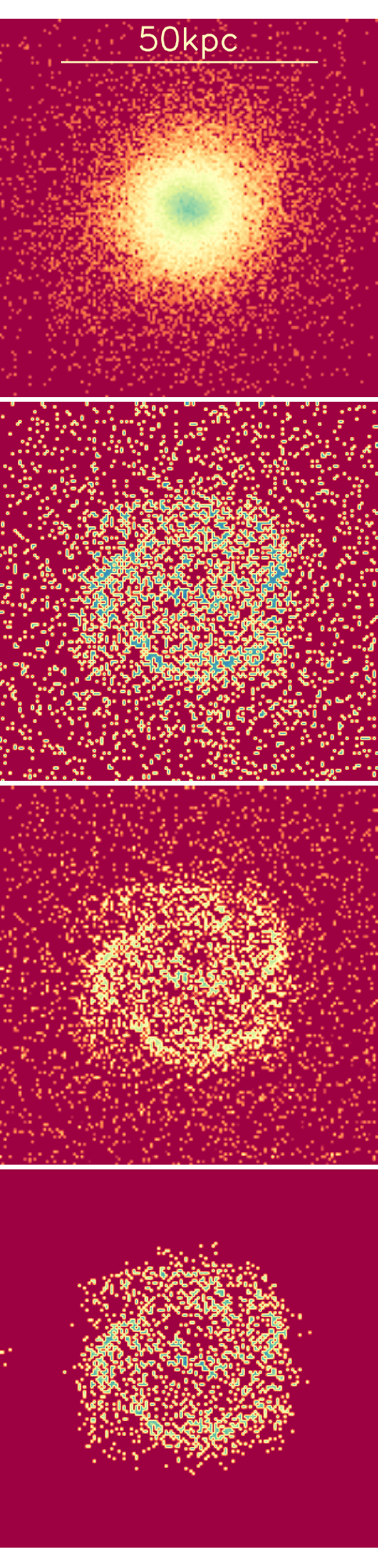}
\includegraphics[width=0.24\textwidth]{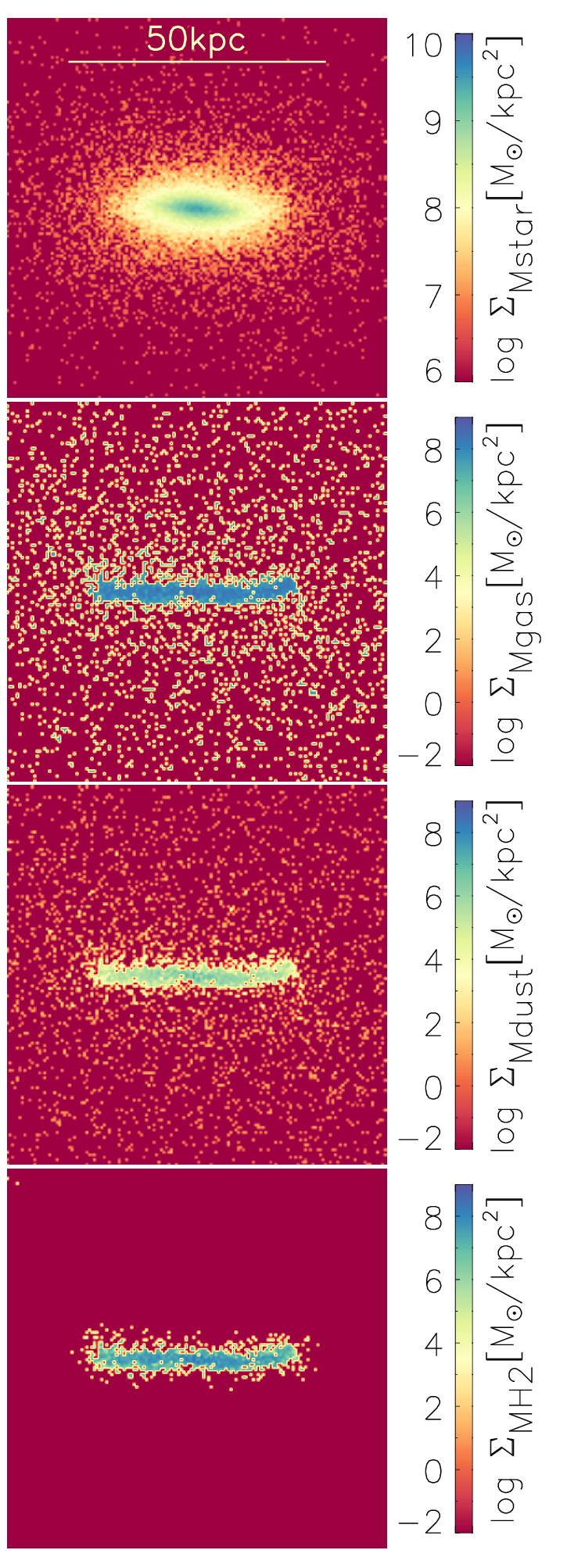}
\hspace{-0.3 truecm}
\includegraphics[width=0.16\textwidth]{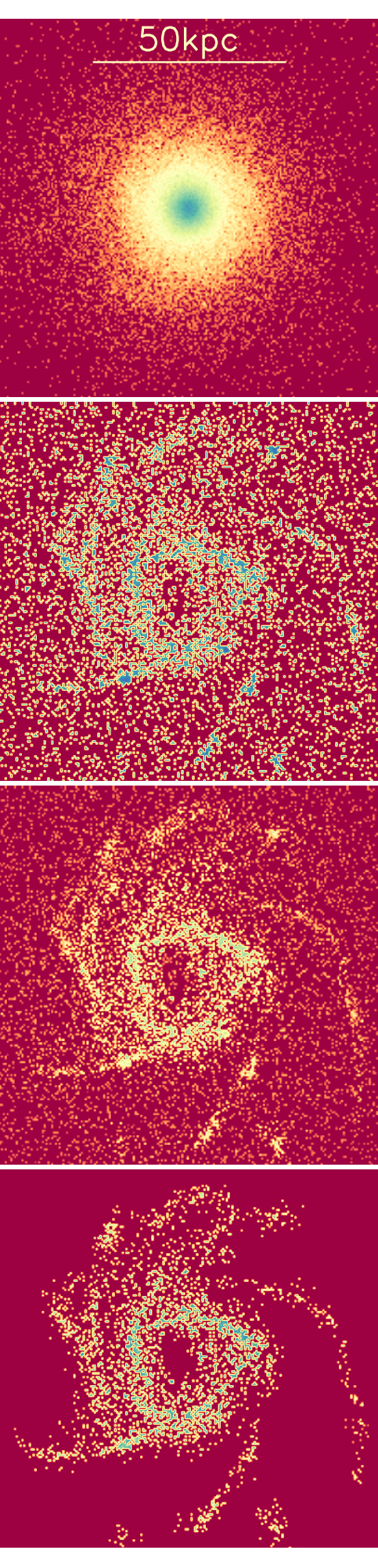}
\includegraphics[width=0.16\textwidth]{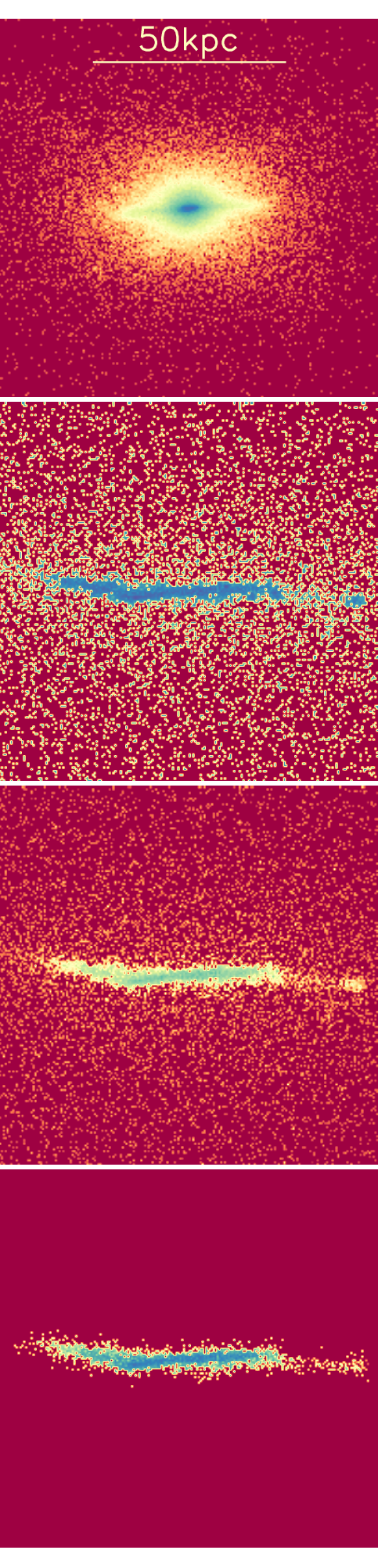}
\caption[width=\textwidth]
{Two examples of disc galaxies from our simulations \rev{at z=0}. The four rows depict the mass surface densities of stars, gas, dust, and ${\rm H_2}$ on pixels of $\sim 0.5 {\rm kpc}$ per side. Columns 1 and 2 illustrate face-on and edge-on views of a galaxy with 
$\rm{M_{200}}=1.5\times10^{12} \msun$, ${\rm M_{star}=5.3\times10^{10} \msun}$, ${\rm M_{gas}=5.0\times10^{9} \msun}$, ${\rm M_{dust}=7.0\times10^{7} \msun}$, and ${\rm M_{H_2}=2.2\times10^{9} \msun}$.
Columns 3 and 4 show the corresponding maps for a galaxy with $\rm{M_{200}}=3.6\times10^{12} \msun$, ${\rm M_{star}=1.2\times10^{11} \msun}$, ${\rm M_{gas}=4.6\times10^{9} \msun}$, ${\rm M_{dust}=6.3\times10^{7} \msun}$, and ${\rm M_{H_2}=2.0\times10^{9} \msun}$.
}
\label{fig:mapMWs}
\end{figure*}

\subsection{Changes with respect to previous MUPPI models}
\label{sec:MUPPImodifications}
\rev{In this work, we introduced a series of modifications to the MUPPI setup described by \cite{valentini23}, which was calibrated using zoom-in simulations of a MW-type galaxy. We remark that the latter work and most recent papers using MUPPI are based on simulations with a mass resolution higher by more than one order of magnitude than that in our cosmological boxes. Resolution dependence of the 'best' MUPPI setup has to be expected at some level, as in any sub-resolution modelling, meaning that the changes discussed below may be at least in part demanded by the differences in resolutions. 
} Together with our treatment of the molecular mass fuel for SF, these modifications significantly improved the agreement with the basic properties of the galaxy population in the cosmological boxes, which are presented in Section \ref{sec:results1}. These adjustments essentially preserve the key achievements noted in earlier zoom-in simulations. \rev{Figure \ref{fig:mapMWs}  presents face-on and edge-on maps of the various baryonic components of two spiral galaxies extracted from the simulations. One galaxy has a baryonic mass of approximately $6 \times 10^{10} \msun$, comparable to that of the Milky Way \citep[e.g.][]{mcmillan2017}, while the other is about twice as massive.} 

\rev{The mentioned modifications to MUPPI are listed below, and Table \ref{tab:muppi_table} provides a summary of them.}
\begin{itemize}
    \item 
    Among the different IMF explored in \cite{valentini19}, we adopt that by \cite{chabrier03}, which 
    is likely the most popular when dealing with observations. 
    \item 
    Following  \cite{valentini19} (readers can refer to their Table 1 and Section 3), we adopt here a lower value for the kinetic wind efficiency $f_{\rm FB, kin}=0.08$, more appropriate for an IMF enriched with massive stars.
    \item \rev{We introduce a dependence of the lifetime of particles in the wind, $t_{\rm wind}$, on the velocity dispersion $\sigma_{DM}$ of neighbouring DM particles.
    Wind particles are designed to sample galactic outflows. 
    A gas particle exits its MP stage after a maximum allowed time given by the \rev{free-fall} time of the cold gas. When a gas particle exits a MP stage, it has a probability $P_{\rm kin}$ (a model parameter) of being kicked and becoming a wind particle for a time interval $t_{\rm wind}$. Wind particles are decoupled from hydro forces during the aforementioned interval $t_{\rm wind}$. Still, they are affected by radiative cooling and receive kinetic energy from neighbouring star-forming gas particles. The latter is used to increase their velocity. For further details, readers can refer to Appendix \ref{app:refMUPPI} and references therein. The dependence of $t_{\rm wind}$ on $\sigma_{DM}$ introduced in the present work reads as follows:}
\begin{equation}
t_{\rm wind} = \left(\frac{\sigma_{0}}{\sigma_{DM}}\right)^4 \frac{t_{\rm ff, c}}{2},
\label{eq:twind}
\end{equation}
    where $\sigma_{0}=50 \, {\rm km \, s}^{-1}$ and $t_{\rm ff, c}$ is the free-fall time of the sub-grid cold phase. 
    For particles with $\sigma_{DM}$ greater (smaller) than $\sigma_{0}$, their time spent in the wind is shortened (prolonged) compared to half their free-fall time. As a result, in less massive haloes, where the typical $\sigma_{DM}$ is smaller, wind particles require more time to return and become eligible for MP once again. This adjustment is crucial to prevent the formation of low-mass galaxies. The value of $\sigma_{0}$ was adjusted to reproduce the Stellar-Halo mass relation at $z=0$.
    %
    In Appendix \ref{app:refMUPPI}, we illustrate how this choice impacts the SFRD, the stellar-halo mass relation (SHMR), and the stellar mass function (SMF) (Fig. \ref{fig:oldsetup1}).
    \item 
    We adopt the low metallicity feedback (LMF) introduced in \cite{valentini23} to mimic feedback by an early population of stars. We assume that star-forming particles with low metallicity contribute to the ISM with kinetic energy with an efficiency larger by a factor of 20 with respect to higher metallicity star-forming particles. In \cite{valentini23}, this factor was 10.
    The impact of this choice is shown in Appendix \ref{app:refMUPPI} (Fig. \ref{fig:oldsetup2}).
    \item  
    The density threshold for entering MP is halved with respect to \cite{valentini23}, so that now it is set to 0.005 $\rm{cm^{-3}}$. This adjustment aids in shifting the peak of the SFRD towards lower redshifts, aligning it with observed determinations. The SF efficiency is also halved and set at $f_* = 0.01$.
    \item 
    The black hole (BH) radiative and feedback efficiencies are set to 0.1.  The halo mass for BH seeding is $\rm{1.3\times10^{11} M_{\odot}}$, with the seeded BH having a mass of $\rm{1.2\times10^{5} M_{\odot}}$. Our simulations do not incorporate radio-mode feedback. As for the pinning of BHs at the centre of the hosting galaxy, our model uses the method detailed in \cite{ragone18}, which avoids non-physical behaviours that can affect less careful BH repositioning    \citep[see also][]{damiano24}.
    \item 
    As for the AGN feedback energy partition between the hot and cold phases of MP particles, we abandon the fiducial solution by \cite{valentini20}, based on an evolving, thought parameterized estimate of the covering factor of cold clouds.  We adopt instead their simpler prescription in which the energy is evenly distributed between the two phases, dubbed both in that paper. The latter prescription yielded qualitatively similar outcomes in their zoom-in MW-like galaxy simulations. However, it lowers the SFRD at low redshifts in our cosmological boxes.
\end{itemize}

\begin{table}[h!]
\caption{Modifications to MUPPI.}
\centering
\renewcommand{\arraystretch}{1.2}
\begin{tabular}{ l c c}
\hline
\textbf{Feature} & \textbf{This Work} & \textbf{Valentini+23} \\ \hline
\text{(1) IMF}      & Chabrier03  &  Kroupa93   \\ 
\text{(2) $f_{\rm FB, kin}$}      & 0.08  & 0.1    \\ 
\text{(3) Time in Wind}      & Eq.\ref{eq:twind}    & $45\ \text{Myr} - t_{\rm ff, c}$   \\ 
\text{(4) LMF}                    & 20    & 10   \\ 
\text{(5) $\rm{n_{MP} [cm^{-3}]}$} & 0.05  & 0.1 \\ 
\text{(6) $f_*$} & 0.01  & 0.02 \\ 
\text{(7) BH: $\epsilon_r$ \& $\epsilon_f$} & 0.1 \,\, 0.1  & 0.1 \,\, 0.01 \\
\text{(8) BH: Halo min mass [$\msun$]} & $1.3\times 10^{11}$  &  $1.8\times 10^{10}$\\ 
\text{(9) AGN Egy Partition} & 1/2  &  CovFact\\ \hline
\end{tabular}
\tablefoot{Differences with respect to the previous MUPPI setup. From top to bottom features are, (1) the initial mass function (Kroupa93 corresponds to \cite{kroupa93}), (2) the kinetic wind efficiency, (3) the maximum lifetime of wind particles $t_{\rm wind}$ prescription, (4) the low metallicity feedback factor, (5) the density threshold for entering MP, (6) the star formation efficiency, (7) the BH radiative and feedback eﬃciencies, (8) the halo mass for BH seeding, and (9) the energy partition between the hot and cold phases  (1/2 means that the energy is evenly distributed between phases and CovFact that the partition uses  an estimate of the covering factor of cold cloud. For more details, readers can refer to Sec. \ref{sec:MUPPImodifications}.}
\label{tab:muppi_table}
\end{table}

\subsection{The dust model}
In this work, the treatment of dust pollution from evolved stars (AGBs and SNae) and subsequent evolution of the grain population follows the model illustrated in our recent works \citep{gjergo18, granato21, parente22}, including the effect of hot gas cooling due to collision with grains. We adopted the choice of parameters calibrated in the last paper. The treatment of dust size distribution and the size dependence of related processes adopts the computationally cheap two-size approximation devised by \cite{hirashita15}. It was derived by comparing with results of one-zone computation, and its satisfactory performances have also been tested against more detailed treatments also by other groups \citep[e.g.][]{aoyama20}.

\section{Treatment of \hmol evolution in MUPPI}
\label{sec:MUPPIH2}
Building on the groundwork of the already implemented dust formation and evolution modelling in MUPPI \citep[see][]{granato21}, the present study takes a step forward by integrating a model for the formation of \hmol on dust grains. This prompts us to introduce an additional equation to the existing set of five differential equations. These equations delineate the evolution of the cold, hot, and "virtual" stellar component, the thermal energy of the hot phase, and the AGN feedback energy acting on the particle \cite[equations 28 to 31 in][]{valentini20}. The new equation describes the mass evolution of the \hmol portion (see below), fully encompassed within the cold phase.

MUPPI  assumes that the molecular gas is the fuel for SF. Therefore, the SFR can be written as
\begin{equation}
    \dot{M}_{\rm{sf}} = f_* \frac{f_{\rm mol} \, M_{\rm c}}{t_{\rm ff}},
\end{equation}
where $f_*$ is an efficiency factor, the \rev{local free-fall time\footnote{in previous MUPPI papers referred to as local dynamical time} $t_{\rm ff}$} refers to the cold gas and $f_{\rm mol}$ is the fraction of the latter in molecular clouds. In previous works, we computed the latter using external prescriptions, either theoretical \citep{Krumholz2009} or phenomenological \citep{Blitz2006} in nature \citep[see][and references therein]{valentini23}.
With our new implementation of molecular hydrogen evolution 
we compute the virtual SFR of MUPPI gas particles as 
\begin{equation}
    \dot{M}_{\rm{sf}}= f_* \frac{M_{\rm H2}}{(1 -Y) t_{\rm ff}}, 
\label{eq:SFlaw}    
\end{equation}
where $Y=0.24$ is the Helium fraction, and $M_{\rm H2}$ is obtained by adding to MUPPI a new differential equation that describes the physical processes involved in the \hmol evolution: 
\begin{eqnarray}
\dot{M}_{\rm{H_2}}&=& \dot{M}_{\rm{formH_2}} - \dot{M}_{\rm{destH_2}} - \dot{M}_{\rm{sf}} - \dot{M}_{\rm{ev}} - {\rm f_{H2}}\ \dot{M}_{\rm{AGN}}.
\label{eq:newmuppi}
\end{eqnarray}
Here the $\dot{M}_{\rm{formH_2}}$ and $\dot{M}_{\rm{destH_2}}$ terms account for the formation of \hmol on dust grains and its destruction by dissociating UV radiation, respectively. 
The terms $\dot{M}_{\rm{ev}}$ and $\dot{M}_{\rm{AGN}}$ were already present in MUPPI, transferring mass from the cold to the hot phase \citep[Eq. 29 in][]{valentini20}. The former represents the "evaporation" of the molecular phase due to stellar winds and is set to $0.1 \dot{M}_{\rm{sf}}$. This destruction term is fully applied to the molecular phase in our treatment. The latter, $\dot{M}_{\rm{AGN}}$, represents the destruction due to the share of AGN feedback energy that couples to the cold phase.  A fraction ${\rm f_{H2}} = M_{\rm H2}/ M_{\rm c}$ of this destruction rate is applied to the molecular phase. A schematic representation of the mass fluxes just described is shown in Fig.\ref{fig:MPparticle}, depicting a MP particle.

\subsection{\hmol Formation on dust grains}
Various theoretical studies, starting with \cite{gould_salpeter_63}\footnote{At that time, the amount of interstellar \hmol had not yet been estimated, but there were already indications of its high abundance.}, have shown that the \hmol formation in the gas-phase within the ISM cannot account for the inferred abundance of \hmol in space. 
According to their estimates, the most important mechanism for forming molecular hydrogen is the encounter of hydrogen atoms on the surface of interstellar grains. This conclusion is currently widely accepted \rev{\citep[see][for a recent review]{wakelam17} and adopted in theoretical works to estimate the \hmol formation rate in dust enriched ISM \citep[e.g.][]{Krumholz2009, Bekki2013, bron14, Tomassetti2015, valdivia16}.} 

\cite{hollenbach71} computed the rate per unit volume of molecular formation on dust grains as
\begin{equation} 
\label{eq:RG}
R_{\rm{G}} = \frac{1}{2} \, \gamma \, \langle v_{\rm{H}} \rangle \, n_{\rm{H}} \, n_{\rm{gr}} \, \langle \sigma_{\rm{gr}} \rangle  ;
\end{equation}
the product of the last four factors gives the rate per unit volume at which HI atoms, with number density $n_{\rm{H}}$ and mean velocities $\langle v_{\rm{H}}\rangle$, collide with dust grains, with number density $n_{\rm{gr}}$ and mean geometric cross-section $\langle \sigma_{\rm{gr}}\rangle$.
The efficiency factor $\gamma$ represents the fraction of HI atoms that combine on the surface of dust grains and detach from it as \hmol molecules, referred to as the recombination coefficient. 

Using equation \ref{eq:RG} and the two-size approximation adopted by our dust evolution model, the total rate of \hmol formation, including both small and large grains, reads as follows:
\begin{equation}
R_{\rm{G}} = \frac{1}{2} \, \gamma \, \langle v_{\rm{H}} \rangle \, n_{\rm{H}} \, \left(n_{\rm{gr,s}} \, \langle \sigma_{\rm{gr,s}} \rangle + n_{\rm{gr,l}} \, \langle \sigma_{\rm{gr,l}} \rangle \right)
\label{eq:RG2}
\end{equation}
where subscripts $\rm{gr,s}$ and $\rm{gr,l}$ stand for small and large grains. Following our previous works, we adopt for them the reference sizes of $0.005 \rm{\mu m}$ and $0.05\rm{\mu m}$, respectively\footnote{To convert from dust grain mass densities to dust grain number densities we consider as representative of material density the figure of $3\,\rm{gr\, cm^{-3}}$ for both graphite and silicate grains.}
\footnote{With the physical properties of grains assumed here, and when the large to small grains mass ratio is $\sim 5$, as usually expected for models reproducing the properties of MW dust \citep[ e.g.][]{granato21}, the estimate performed by \cite{hollenbach71} from Eq. \ref{eq:RG} to their equation 3 yields a rate of \hmol formation $5 \times 10^{-17} n_{\rm H} n \, \rm{cm}^{-3} s^{-1}$. This is relatively close to the value $3.5 \times 10^{-17} n_{\rm H} n \,  \rm{cm}^{-3} s^{-1}$, adopted at solar metallicity by previous simulations
\citep[e.g.][]{Pelupessy2006, Gnedin2009, Christensen2012}. We remember that in those works, dust evolution was not treated. 
The dust-to-gas ratio was estimated by simply scaling it linearly with metallicity}. Finally, by setting
$\gamma \langle v_{\rm{H}}\rangle = 5\times10^4\, \rm{cm\,sec^{-1}}$ \citep{hollenbach71}, we  can estimate the term for \hmol formation in Eq. \ref{eq:newmuppi} as
\begin{equation}
\dot{M}_{\rm{formH_2}} = C_{\rho}\, 2\, \rm{m_p} \, R_{\rm{G}} \,V_c 
\label{eq:hmolform}
\end{equation}
where $\rm{m_p}$ is the proton mass, $V_c$ is the cold phase volume\footnote{The cold phase volume is obtained from the condition of pressure equilibrium between the hot and cold phases, see Section 2.1 in \cite{murante10}} and the clumping factor $C_{\rho}=10$ is intended to boost the density of the cold phase to account for the unresolved molecular cloud densities in our model.

Since gas particles initiate their evolution without dust, the formation of \hmol, and consequently SF, would never begin without some initial "seeding" of \hmol. To address this issue, whenever a particle enters MP, and its D/G ratio is lower than 0.03 (D/G)$_\odot$, with (D/G)$_\odot$=0.01, SF is conducted using the empirical prescription from \cite{Blitz2006}, consistent with the previous version of MUPPI. We verified that our results are stable against changes in these thresholds by a factor of at least two.


\begin{figure}
\centering
\includegraphics[width=0.45\textwidth]{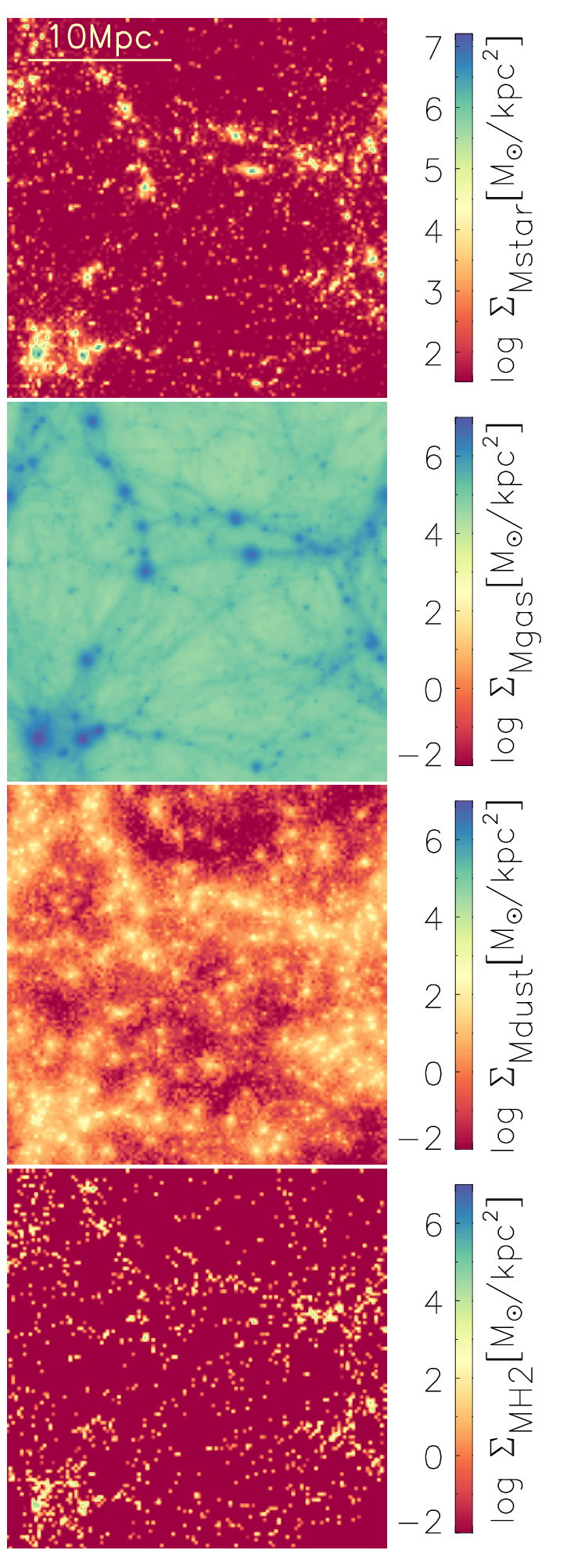}
\caption[width=\textwidth]
{Stellar, gas, dust, and \hmol mass density maps in one of the $(26~\rm Mpc)^3$ simulation box computed in pixels of 200x200 kpc$^2$. }
\label{fig:map}
\end{figure}

\subsection{The photo-dissociation rate}
Photons from recently formed stars in the LW region ($\sim 940-1070 \textup{~\AA}$) are particularly harmful to \hmol. 
The rate at which \hmol dissociates is estimated following \cite{abel1997}:
\begin{equation}
    k_{\rm LW} = C_{\rm LW} \; 4 \pi \, J_{\nu ,{\rm LW} }
\end{equation}
where $J_{\nu,{\rm LW}}$ denotes the specific radiative intensity averaged over angles at
$h \nu = 12.87 \rm{eV}$, 
and $C_{\rm LW} 
= 1.1 \times 10^{8} \,
\mbox{erg}^{-1}
\mbox{cm}^{2} \,\mbox{sr} \, \mbox{Hz} $.


However, we point out that several model computations \citep[e.g.][]{Gnedin2009, maclow12, Krumholz2009, Christensen2012} suggest that this process may be subdominant in setting the molecular gas fraction at galactic scales in situations where \hmol production is dominated by dust surface catalysis, that is in the conditions to which our modelling applies. This conclusion derives from the high efficiency of \hmol self-shielding and dust shielding, effectively protecting most of the volume occupied by $\rm H_2$. 
Nevertheless, we implemented an approximate treatment of LW \hmol dissociation to have a first-order mean to estimate its possible importance.

Rigorously accounting for this \hmol destruction effect in cosmological simulations is currently unfeasible. It would require on-the-fly radiative transfer computation in a dusty medium, including the \hmol self-absorption. However, besides the prohibitive numerical cost, the distributions of the relevant emitting sources (stars younger than 5-10 Myrs) and the dense absorbing medium are unresolved in such simulations. 
Therefore, previous works dealing with this aspect in a similar context \citep[e.g.][]{Gnedin2009,Christensen2012} adopted simplified treatments.  Our approach, described in the following, focuses on modelling the phenomena at unresolved scales (namely within MP gas particles), where we believe the bulk of the effect arises. We establish a simple connection between the dissociation rate of \hmol and the recent SFR within an MP particle.

The starting point is to notice that most LW photons come from stars still embedded or relatively close to (Giant) MC that are optically thick at these short wavelengths. We pre-compute the specific intrinsic (meaning before any absorption effect) luminosity for a constant SFR with GRASIL \citep{silva98,granato00}, at the LW frequency, as a function of age $t$.
For the Chabrier IMF adopted in this work, the result is well described by the following polynomial:
\begin{equation}
\log \frac{L_{\nu,{\rm LW}}[{\mbox{erg s}^{-1} \mbox{Hz}^{-1}}]}{\mbox{SFR}[\mbox{M}_\odot \mbox{yr}^{-1}]} = 0.126\, x^3 - 0.774\, x^2 +1.573\, x + 27.03
    \label{eq:LWrad}
\end{equation}
where $x= \log(t[Myr])$. This expression reproduces almost exactly the population synthesis result for $t$ ranging from 0.3 to 300 Myr and stellar metallicity close to $0.3 \, Z_\odot$. After 100 Myr the LW luminosity saturates to the 100 Myr value, with a contribution from stars older than 20-30 Myr lower than 20\%. The metallicity dependence is weak enough to ensure accuracy within 20\% between $0.001$~$Z_\odot$ to $3$~$Z_\odot$. 

We use Eq. \ref{eq:LWrad} to compute, at a given time the LW radiation emitted by stars (usually virtual stars before the spawning of a new stellar particle) born in the MP particles during the previous timesteps of the MP cycle, by finite difference. However, dust attenuates LW radiation efficiently, substantially protecting \hmol from LW photo-dissociation. Moreover, also \hmol self-shielding has a decisive protective effect. 
We refer the reader to the Appendix \ref{app:shield} for details on our estimate of the shielding factor. Here, we point out the primary outcome of this estimate. 
\rev{If the physical properties of the MC in the Milky Way \citep{miville17} are representative of the general population of MC, we expect the term $\dot{M}_{\rm{destH_2}}$ in Eq.\ref{eq:newmuppi} to play a negligible role. This conclusion is further strengthened if, as observations suggest \citep{dessauges19}, MC in galaxies at higher redshift and/or with higher star formation activity exhibit higher column densities.}
Even by artificially enhancing the LW radiation field given by Eq. \ref{eq:LWrad} by a factor of 10, the results are almost indistinguishable from those obtained by completely neglecting the LW destruction, indicating that the conclusion is robust.  

\begin{figure}
\centering
\includegraphics[width=0.5\textwidth]{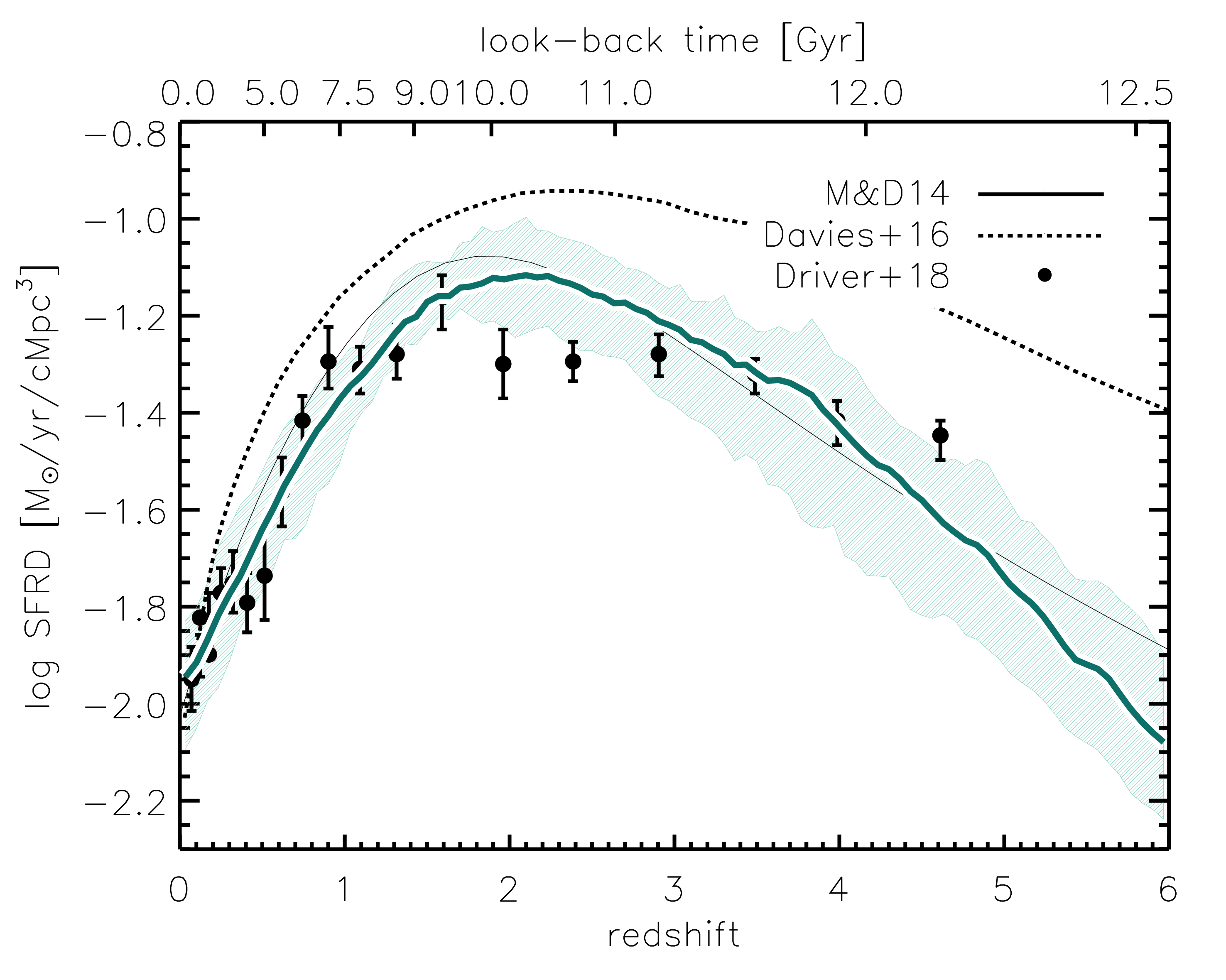}
\caption[width=\textwidth]
{Average cosmic SFRD calculated from five $(26~\rm{Mpc})^3$ simulation boxes, with the shaded region indicating the full dispersion of the curves. We compare with some determinations from literature: solid and dashed black lines represent the fits to the cosmic SFRD obtained by \cite{Madau2014} and \cite{davies16} respectively, while black circles show the determination of \cite{driver18}. 
Following their recommendation, the data from \cite{Madau2014} have been scaled by a factor of 0.63 to convert from a \cite{salpeter95} to a \cite{chabrier03} IMF. The remaining data points require no further conversion.}
\label{fig:sfrd}
\end{figure}

\begin{figure*}
\centering
\includegraphics[width=\textwidth]{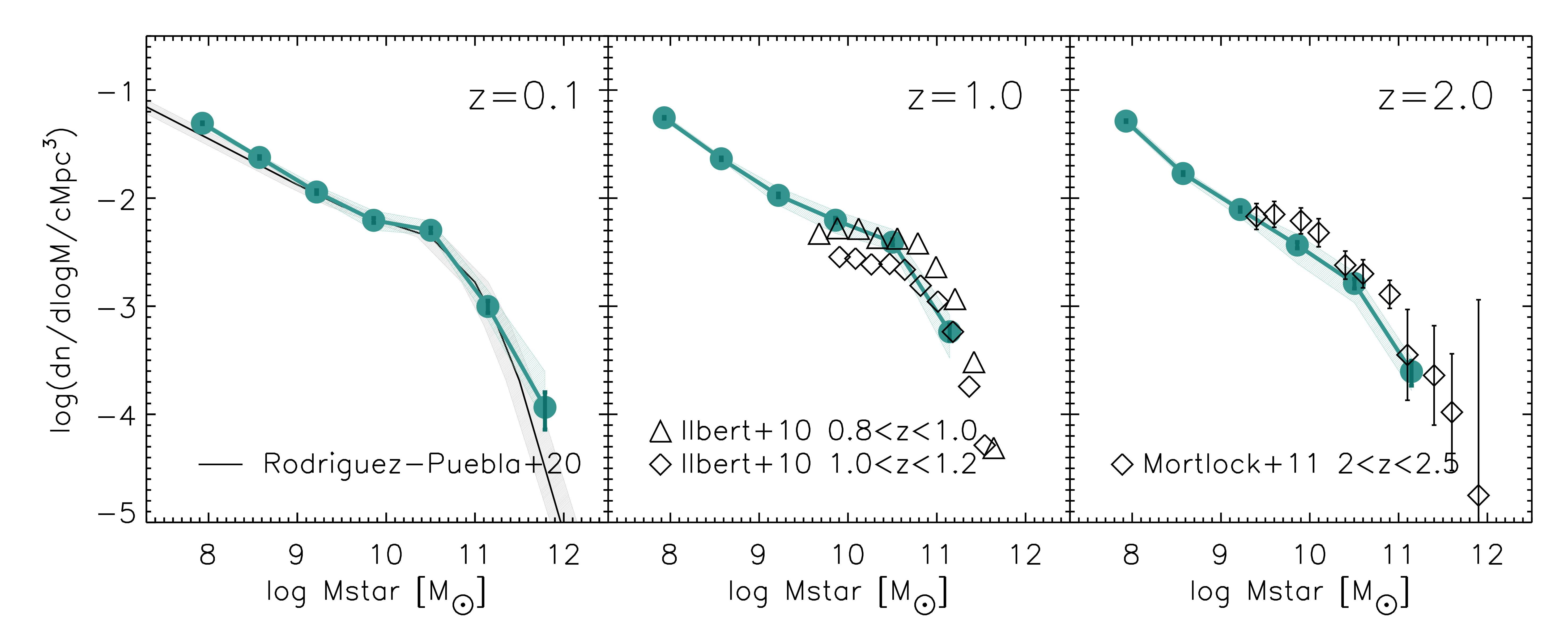}
\caption[width=\textwidth]
{Stellar mass function is shown by the green curves at $z=0.1$, $z=1$, and $z=2$, based on the five simulated boxes in our fiducial model. The error bars represent Poissonian errors for each stellar mass bin, typically smaller than the size of the point. \rev{The shaded green region encompasses the full dispersion of the individual SMFs corresponding to each of the five simulation boxes.} We compare our model at $z=0.1$ with the best-fit model of \cite{rodriguez-puebla20}; the grey area is an estimate of systematic errors which captures variations due to differences in the mass-to-light ratios and mass definitions (photometry) of the observational samples in that work. Simulations at $z=1$ are confronted with the data from \cite{ilbert10} at two redshift ranges and at $z=2$ with the data in \cite{Mortlock2011}.
}
\label{fig:smf}
\end{figure*}

\begin{figure*}
\centering
\includegraphics[width=\textwidth]{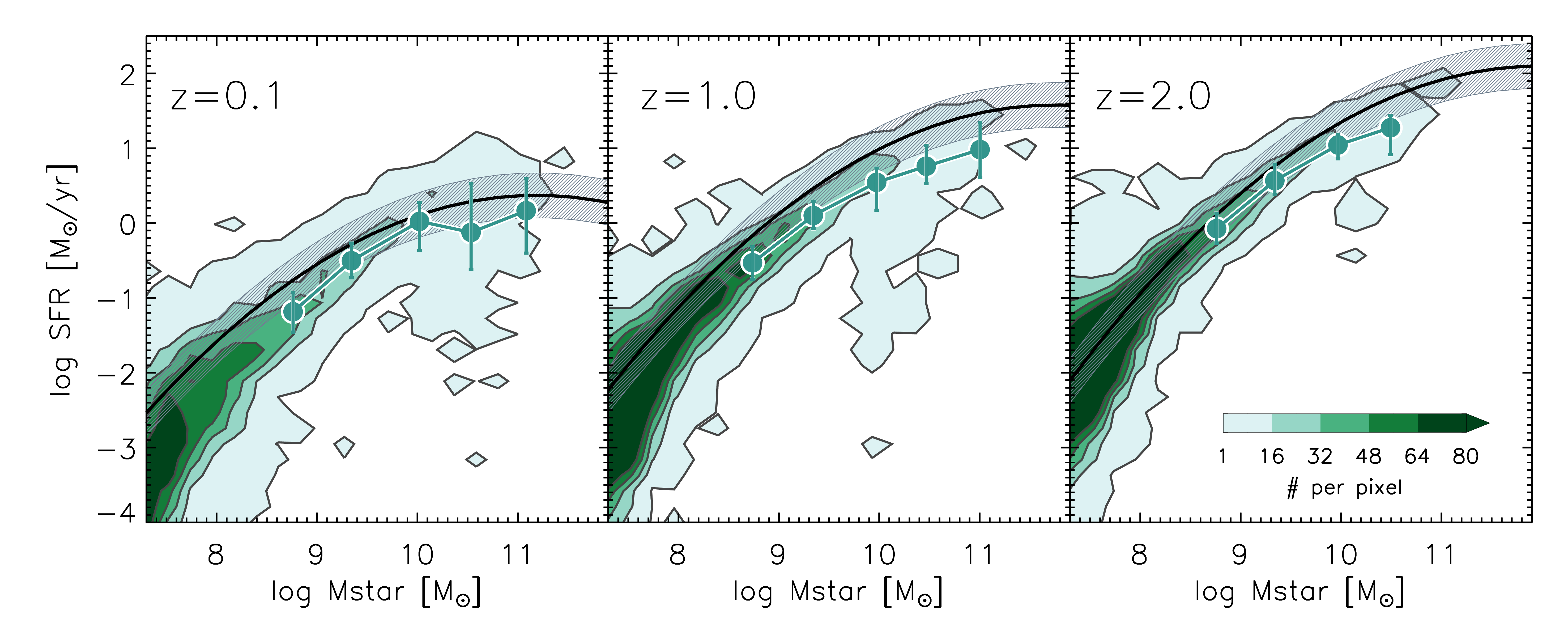}
\caption[width=\textwidth]
{Distribution of our simulated galaxies in the stellar mass–SFR plane is shown by the contours. Solid black lines are the MS determinations for observed galaxies presented in \cite{popesso23}, resulting from an extensive homogenized compilation of studies published since 2014 (their eq. 10). The shaded grey band visualize the typical dispersion of determinations used to get their fit. The green curves connecting the points represent the median SFR per mass bin for a sub-sample of simulated galaxies selected using the same criteria as in \cite{popesso23}, namely $\rm{8.5 < log(Mstar/\msun) < 11.5}$ and $\rm{0.01 < SFR/(\msun yr^{-1}) < 500}$. Bars delineate the 25-75\% percentiles.
}
\label{fig:mainseq}
\end{figure*}

\section{Model comparison with observational data}
\label{sec:results1}
To validate our model, this section contrasts key statistical characteristics of simulated galaxies across our five boxes, introduced in Section \ref{sec:inicond}, with existing observational data.
We adjusted the model parameters to find the best match with the observed cosmic SFRD, the stellar mass function, and the stellar mass-halo mass function at $z=0$. Consequently, the remaining results presented here can be considered predictions of our model. A visual rendition of the outcome of our model at redshift $z=0$ can be seen in Fig. \ref{fig:map}, which illustrates the density maps for the various components within one of our simulated volumes.

To identify galaxies, we employ \textsc{Subfind} \citep{Springel2001, Dolag2009}, considering objects with stellar masses larger than $3 \times 10^7 \msun$, which corresponds to an average of 25 stellar particles. 
This low number of minimum resolved stellar elements should be sufficient to reasonably define the basic properties of galaxies considered in the present work, i.e.\ masses of the various components. 
The integrated masses and SFR in this study are evaluated within twice the half-stellar mass radius. These computations focus solely on particles that are linked to a specific subhalo as identified by \textsc{Subfind}.

We begin by examining properties related to the stellar and dust components. These comparisons were feasible even before integrating the \hmol evolution into our modelling \citep[see][]{parente22}. 
However, we have now achieved notable improvements in aligning with the data, thanks to the modifications we introduced in MUPPI and, to some extent, the inclusion of the \hmol modelling.

\subsection{Stellar properties of the galaxy population}
\label{sec:resstars}
In this subsection, we test our simulations against the main observed statistical properties of the stellar component of galaxies, namely the cosmic SFRD, the SMF, the main sequence of star-forming galaxies (MS), and the SHMR, \rev{the latter three at $z=0$, 1, and 2. We reiterate that the SFRD, SMF, and the SHMR at $z=0$ have been used to calibrate the model parameters, and therefore they are not model predictions. Nevertheless, it is not trivial that the model can reproduce those observables with a proper choice of free parameters that are fixed for the rest of our study.}

\subsubsection{The star formation rate density}
The green curve in Fig. \ref{fig:sfrd} shows the average SFRD from the five simulated boxes and the full dispersion of the five individual curves. Our model utilizes a \cite{chabrier03} IMF, and the displayed data have been reported to this same IMF whenever necessary.
The agreement with observational determinations is quite good, comfortably fitting within the existing range of variations among them. This improvement is pronounced compared to the previous version of MUPPI model \citep[e.g.][]{parente22}, which relied on estimating molecular mass through the empirical law proposed by \cite{Blitz2006}, alongside a setup calibrated only on zoom-in simulations of MW-like galaxies (see Section \ref{sec:MUPPImodifications}).


\subsubsection{The stellar mass function}
The observational galaxy SMF at redshift $z=0.1,\,1.0$, and 2, shown in Fig. \ref{fig:smf}, is also well recovered by our model, up to the stellar mass statistically accessible by the simulated volume. We compare our model at $z=0.1$ with the best fit obtained in \cite{rodriguez-puebla20} for their compilation of data from different works. The shaded grey region represents systematic errors, encompassing variations arising from differences in mass-to-light ratios and mass definitions (photometry) within the observed comparison samples utilized in that study. 
A similar good agreement is found when contrasting our findings with data from \cite{ilbert10} and \cite{Mortlock2011} for $z>0$. However, our SMF does not encompass the higher mass ranges observed in their data. This limitation stems from this study's relatively small simulation volumes, (26 Mpc)$^3$.

\subsubsection{The main sequence}
\label{sec:MS}
The Mstar vs. SFR relation is illustrated in Fig. \ref{fig:mainseq}, with green contours depicting our complete sample of galaxies. We contrast our predicted MS of star-forming galaxies, represented by the green curve connecting the data points, with the observational findings outlined in \cite{popesso23}, which summarizes an extensive homogenized compilation of studies published since 2014. 
Our results refer to a sub-sample of simulated galaxies selected using the same criteria as in the latter paper, namely $ \rm{8.5 < log(Mstar/\msun) < 11.5}$ and $\rm{0.01 < SFR/(\msun yr^{-1}) < 500}$.
The normalization is generally reasonably well reproduced,  with a tendency to slightly under-predict the observed MS at a given stellar mass. 

Across the various determinations included in the \cite{popesso23} compilation, there is a variation in the redshift and stellar mass at which the MS bends. Nevertheless, after converting all observations to a common calibration, they find that the main sequence displays a curvature towards higher stellar masses at all redshifts. 
Notably, our findings indicate the continuous curvature of this relationship in the log-log plane, consistent with the observational estimate. In contrast, other cosmological simulations, including Illustris, Eagle, Mufasa, IllustrisTNG, and Simba, predict a linear correlation, at least up to high masses \citep[e.g.,][and references therein]{donnari19,hahn19,Dave2019}.
In our model, a similar bending is also present in the  MH2-Mstar relation (Section \ref{sec:resh2}). Considering that \hmol is the fuel for SF, the latter implies that the bending of the MS is mainly driven by a relative lack of \hmol in the most massive galaxies.

Finally, we note that our model also predicts a population of passive galaxies, which is increasingly clear with decreasing redshift. At $z=0.1$ the population is located at ${\rm log \, Mstar/M_\odot \sim 10.5}$ and ${\rm SFR} \sim 0.1 \, \msun/{\rm yr}$, broadly consistent with observations \citep[e.g.,][]{renzini18}.

\subsubsection{The stellar-halo mass relation}
\begin{figure}
\centering
\includegraphics[width=0.5\textwidth]{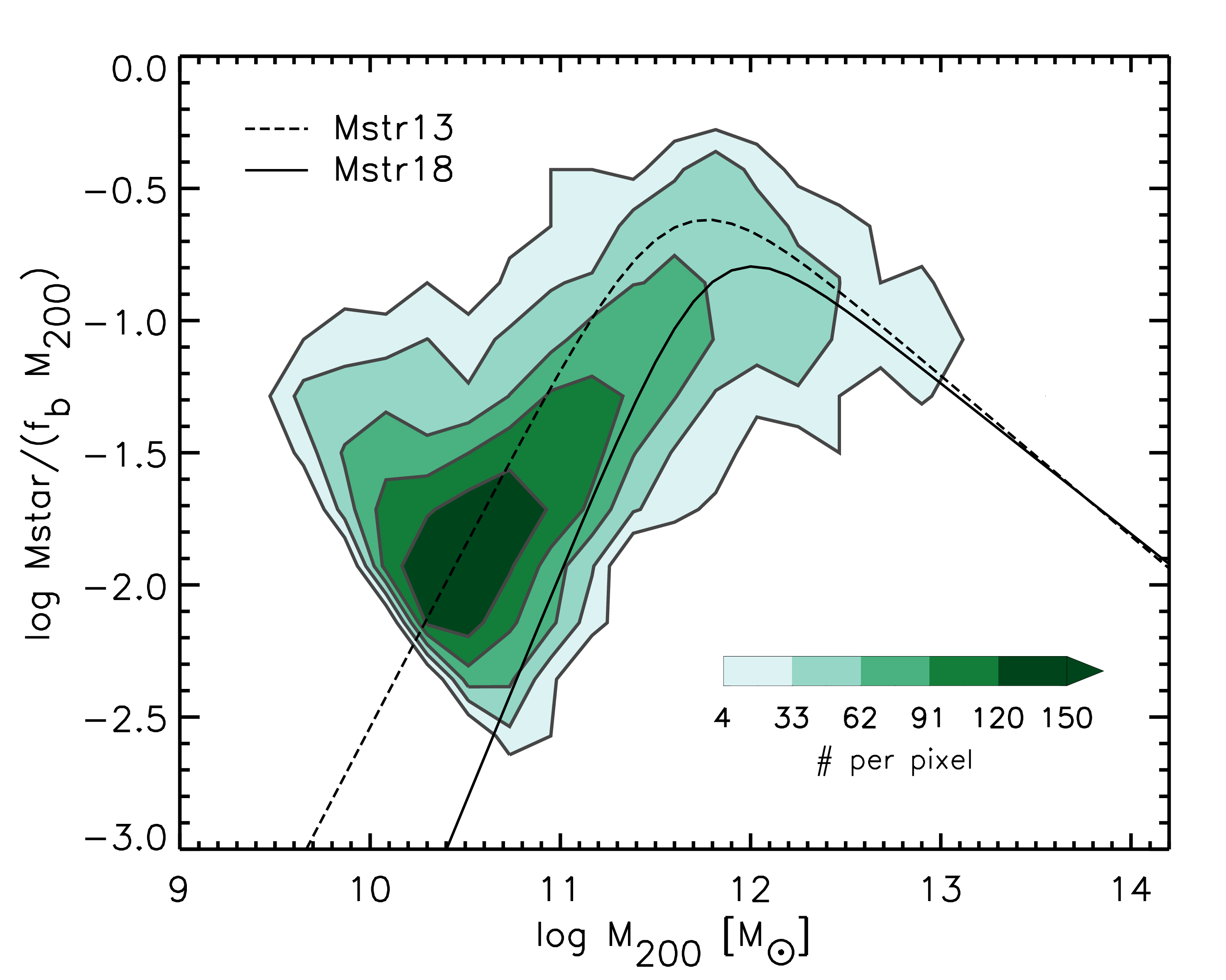}
\caption[width=\textwidth]
{Stellar mass-halo mass relation at $z=0$ is assessed by considering galaxies across all our five simulated boxes, $\rm{f_b}=0.157$ is the baryon fraction $\Omega_{\rm b}/\Omega_{\rm m}$. The stellar mass is computed within $0.1 \rtwo$. The colour contours delineate grids containing a specific number of galaxies as indicated in the colour bar, utilizing 0.2 and 0.2 dex-wide bins along the x and y axes, respectively. We compare our results with abundance matching estimate by \cite{moster13}, and with the empirical model by \cite{moster18}.}
\label{fig:smhm}
\end{figure}

Figure \ref{fig:smhm} shows the $z=0$ relationship between the $\mtwo$ halo mass and the stellar mass of central galaxies, delineated by the baryon conversion efficiency $\rm{\epsilon=Mstar/(f_b\ M_{200}})$, with $\rm {f_b=\Omega_ b/\Omega_m}$. We compare our results with the abundance matching estimate by \cite{moster13} and with the empirical model by \cite{moster18}. The two latter determinations are strongly data-driven, particularly the abundance matching determination, which agrees with our result.
Although our simulation volume does not allow us to extensively test our model in massive ($\gtrsim 10^{13}\,\msun$) halos, simulated galaxies feature a decreasing $\rm{\epsilon}$ at $\rm{M_{200}\gtrsim 10^{12}\,M_\odot}$, in agreement with the reported data-based determinations.

\subsection{Dust mass function}
\begin{figure*}
\centering
\includegraphics[width=\textwidth]{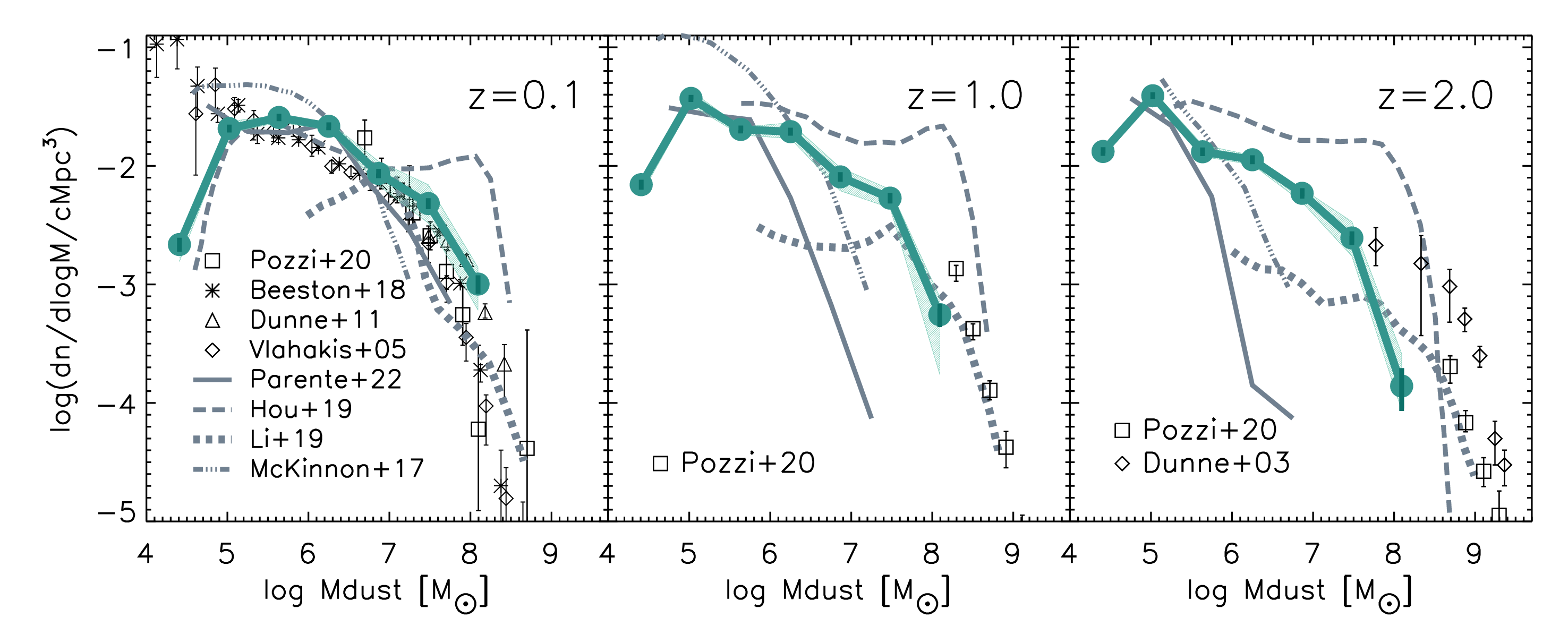}
\caption[width=\textwidth]
{Dust mass function at $z=0.1$, $z=1$, and $z=2$ considering galaxies in the five simulated boxes are represented by green dots connected by solid lines. The error bars represent the Poissonian error for each dust mass bin. \rev{The shaded green region encompasses the full dispersion of the individual DMF corresponding to each of the five simulation boxes.} The decline of the DMF at low masses is due to the imposed stellar mass resolution limit. We compare our results with observational data, represented by symbols and simulations delineated by lines. 
Observations at $z=0.1$ correspond to
\citet{Pozzi2020}, \citet{Beeston2018}, \citet{Dunne2011}, and \citet{Vlahakis2005}, at $z=1$ \citet{Pozzi2020} and at $z=2$ \citet{Pozzi2020} and \citet{Dunne2003}. Simulations correspond to the works of \citet{parente22}, 
\citet{hou2019}, 
\citet{li2019}, and
\citet{mckinnon2017}. 
}
\label{fig:dmf}
\end{figure*}

Fig. \ref{fig:dmf} shows the predicted dust mass function (DMF) at different redshifts and compares it with observational data (black symbols) and other simulations (grey lines). 
In general, previous hydrodynamic simulations \citep{mckinnon2017,hou2019,li2019} had difficulties in reproducing the dust mass function simultaneously at low and high redshift. 
Our simulation outcomes are in good agreement with local observations.  
However, accurately probing redshifts of $z=1$ and $z=2$ presents some challenges. Observations lack coverage of the range of small dust masses that our simulations sample. Conversely, the limited volume of our simulations could lead to an inability to capture the high masses observed in real data, consequently resulting in an under-prediction of the DMF at the highest dust mass bin.

We reiterate the significant improvement over our previous model, as discussed in \cite{parente22}, represented by the solid grey line in the figure. Importantly, this enhancement was achieved by keeping the dust model unchanged while modifying aspects of the galaxy evolution model (see Section \ref{sec:MUPPImodifications}) and including the \hmol modelling.

\begin{figure}
\centering
\includegraphics[width=\columnwidth]{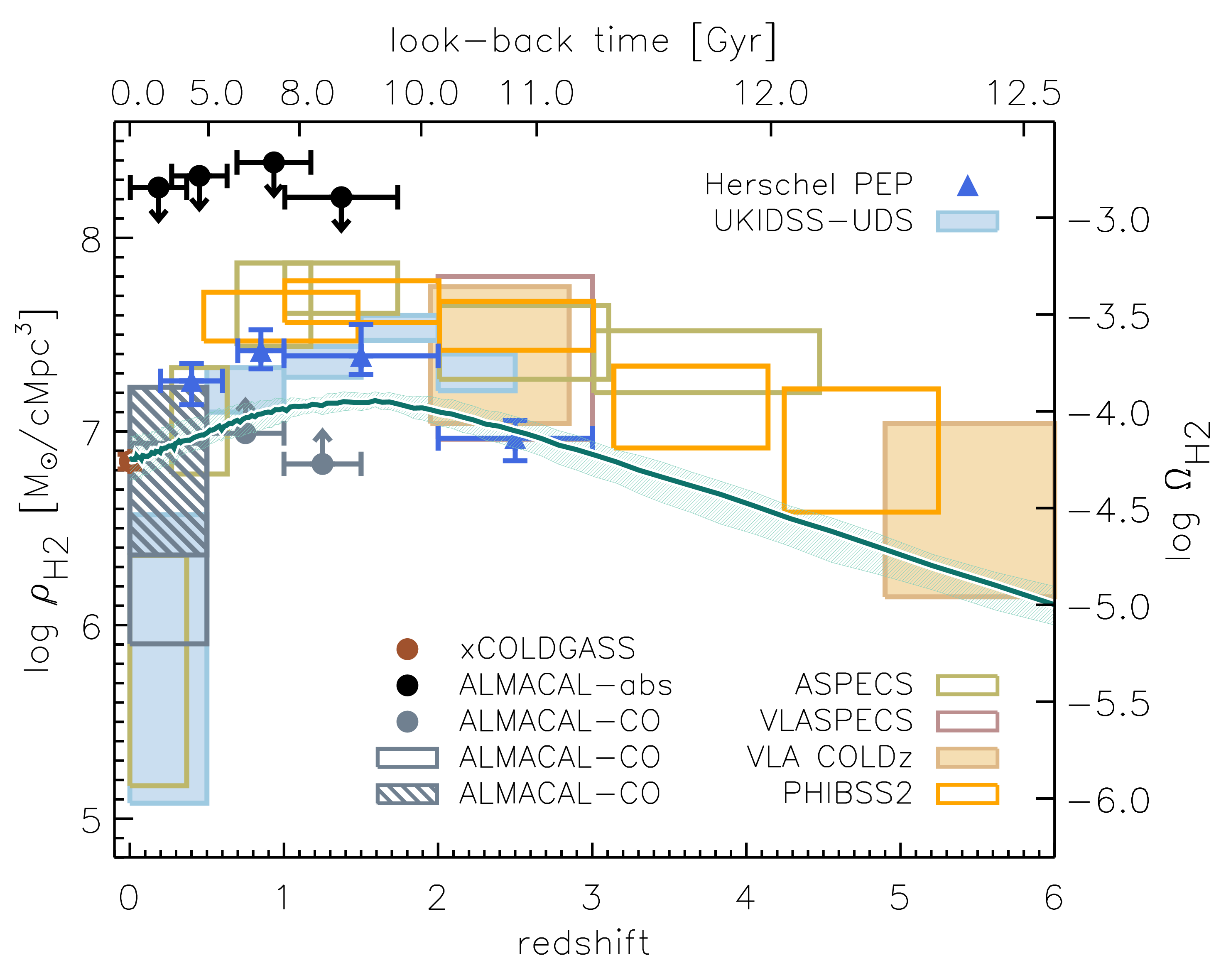}
\caption
{Mean cosmic evolution of the molecular hydrogen density considering the five simulated boxes. The shaded region encloses the full dispersion of the five curves. The right-hand axis shows $\Omega_{\rm{H2}}$, the \hmol mass density normalised to the present-day critical density. Observational constraints are 
ASPECS (ALMA Spectroscopic Survey in the Hubble Ultra Deep Field, \cite{decarli20});
VLASPECS (which uses the NSF’s Karl G. Jansky Very Large Array, VLA, covering part of the ASPECS footprint, \cite{riechers2020}); 
COLDz (VLA CO Luminosity Density at High Redshift in the COSMOS and GOODS-North fields combined; \cite{riechers2019});
PHIBBS2 (Plateau de Bure High-z Blue Sequence Survey 2, which uses the NOrthern Extended Millimeter Array -NOEMA- to observe  field sources, \cite{lenkic2020}); 
xCOLD GASS (eXtended CO Legacy Database for the Galaxy All Sky Survey, \cite{fletcher2021}; 
ALMACAL-abs (blind search of CO absorbers against the ALMA Calibrator archive sources, \cite{klitsch2019}); 
ALMACAL-CO (CO emission-line survey \cite{hamanowicz2023});
Herschel PEP (PACS Evolutionary Probes Herschel survey in the GOODS-N -S fields, \cite{berta2013}); and
UKIDS-UDS (UK InfraRed Telescope -UKIRT- Infrared Deep Sky Survey in the Ultra-Deep Survey, \cite{garratt2021}).
Data listed in the top right legend do not use direct measurement of CO emission to estimate molecular masses.
}
\label{fig:h2d}
\end{figure}

\subsection{\hmol in model galaxies}
\label{sec:resh2}
In the present sub-section, we discuss quantities related to the \hmol content of galaxies, namely the \hmol cosmic density, the \hmol mass function, and the stellar-\hmol mass relation at various redshifts. 

Fig. \ref{fig:h2d} illustrates the evolution of the cosmic \hmol content, showcasing the molecular gas density $\rho_{\rm{H2}}$ for our model alongside constraints derived from observations. These constraints comprise an update of the compilation presented in \cite{maio22}. Despite being by far the most abundant molecule, the \hmol content of galaxies cannot be estimated directly due to its lack of emission at the cold temperatures at which most of its mass resides in MC. The reported estimates are mainly based on CO emission, assuming some conversion factor from CO to $\rm {H_2}$, or on IR emission from dust, assuming a D/G ratio. Both methods suffer from large systematic uncertainties, $\sim 2$, widely debated in the literature and likely increasing with redshift \citep{Tacconi2020}.
Nevertheless, CO-based, dust-based methods and our model predict a molecular gas cosmic density peak at z$\sim$1.5, and align with the broad trend of decreasing over the last $\sim 9$ billion years. The evolution of $\rho_{\rm{H2}}$ in our simulated box is somewhat milder than the observational data suggested. However, the choice of the CO-to-$\rm{H_2}$ conversion factor influences the latter. Indeed, \cite{decarli19} find, using the ASPECS-Pilot survey, that the decline of $\rho_{\rm{H2}}$ by a factor of approximately 6 from its peak at $z\sim 1.5$ to the present would reduce to a factor of 3 if they had used a conversion factor of 2$\msun \rm{(K\ km s^-1 pc^{-2} )^{-1}}$ instead of their favoured choice 3.6 $\msun \rm{(K\ km s^-1 pc^{-2} )^{-1}}$ for galaxies at z>1.

\rev{While taking the various conversion factors adopted by the data in Fig. \ref{fig:h2d} at face value, our prediction, though exhibiting the same qualitative behaviour, underestimates the measurements at $z \sim$ 1 to 2 by a factor $\sim 2$ and lies in the lower limit of observational constraints for higher redshift. We cannot exclude that at very high redshifts when the dust cosmic abundance was likely much lower than it is now, alternative \hmol formation channels, other than dust catalysis, could improve the match with the data. These processes are not included in our computations.
However, this is very unlikely to occur in the redshift range 1 to 2, where the tension with current observational estimates is more evident. In fact, in this redshift regime, cosmic dust abundance is generally estimated to be even higher than in the present-day Universe \citep[see e.g. Fig. 10 in][]{parente23}.}

\begin{figure*}
\centering
\includegraphics[width=\textwidth]{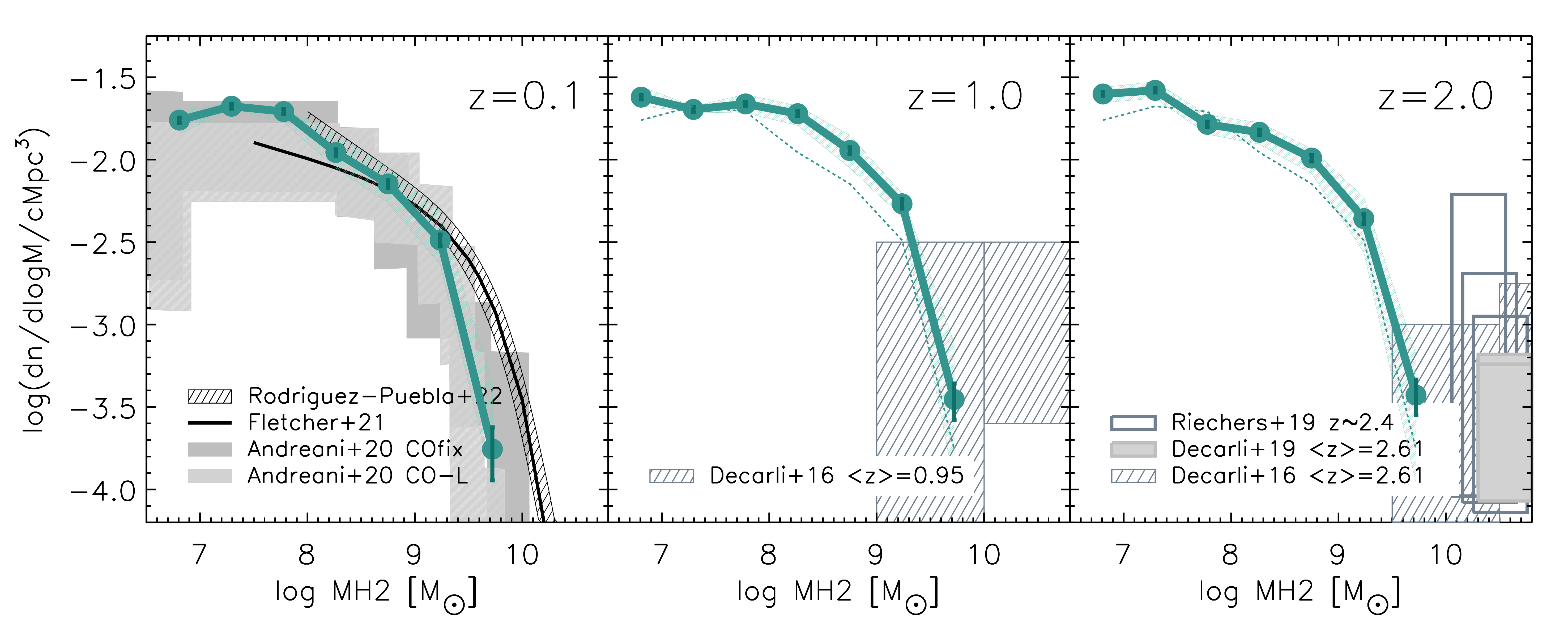}
\caption[width=\textwidth]
{\hmol mass function at $z=0.1$, 1.0, and $z=2.0$. The error bars represent Poissonian errors for each stellar mass bin, typically smaller than the size of the point.
\rev{The shaded green region encompasses the full dispersion of the individual $\rm{H_2}$MFs corresponding to each of the five simulation boxes.}
In the local Universe, we report for comparison the \hmol mass function obtained by \cite{andreani2020}, who use a fixed (COfix) or a luminosity dependent (CO-L) CO-to-$\rm{H_2}$ conversion factor and the best-fit model of \cite{rodriguez-puebla20} and \cite{fletcher2021}.
At higher redshifts, the boxes show constraints from  \cite{decarli16} (ALMA ASPECS-Pilot survey), \cite{decarli19} (ALMA ASPECS-LP survey), and \cite{riechers2019} (COLDz survey). The H2MF at redshift $z=0$ is also shown in the middle and right panels using a thin dotted line to facilitate comparison.
}
\label{fig:h2mf}
\end{figure*}

We also analyse the \hmol mass function (H2MF) at different redshifts. In Fig. \ref{fig:h2mf}, the H2MF in the Local Universe is compared with the determination of \cite{rodriguez-puebla20}, \cite{fletcher2021}, and \cite{andreani2020}. Our predictions closely reproduce the results of \cite{andreani2020}, particularly those adopting the CO-to-$\rm{H_2}$ conversion factor depending on luminosity.
At $z=1$ and $z=2$, the observational validation of the H2MF is limited to the highest masses. An inspection of the figure reveals that at $z=1$ the model H2MF falls, in the most massive bin, within the observed constraints outlined in \cite{decarli16} whose determination corresponds to the redshift bin $z=[0.695-1.174]$ with $<z>=0.95$ 
Similarly, at redshift $z=2$, our H2MF at the highest mass bin falls within the constraints by \cite{decarli16}, despite the latter corresponding to a higher redshift bin. The other two observational constraints depicted in the plot are derived also using galaxy samples at higher redshifts than our determination: \cite{decarli19}, which mirrors the methodology of \cite{decarli16} but in a larger volume, and \cite{riechers2019}.

We obtain little evolution in the H2MF from $z \sim 2$ to the present. This becomes apparent when comparing the determinations of the H2MF at redshifts $z=1$ and $z=2$ with the thin dotted line in Fig \ref{fig:h2mf}, representing the H2MF at redshift $z=0$. This finding contrasts with, for example, \cite{riechers2019} and \cite{decarli19}, who report a shift of the characteristic CO luminosity towards lower values of one order of magnitude from z$\sim$ 2 to 0 (an equal shift in the H2MF for a constant CO-to-$\rm{H_2}$ conversion factor).

\begin{figure*}
\centering
\includegraphics[width=\textwidth]{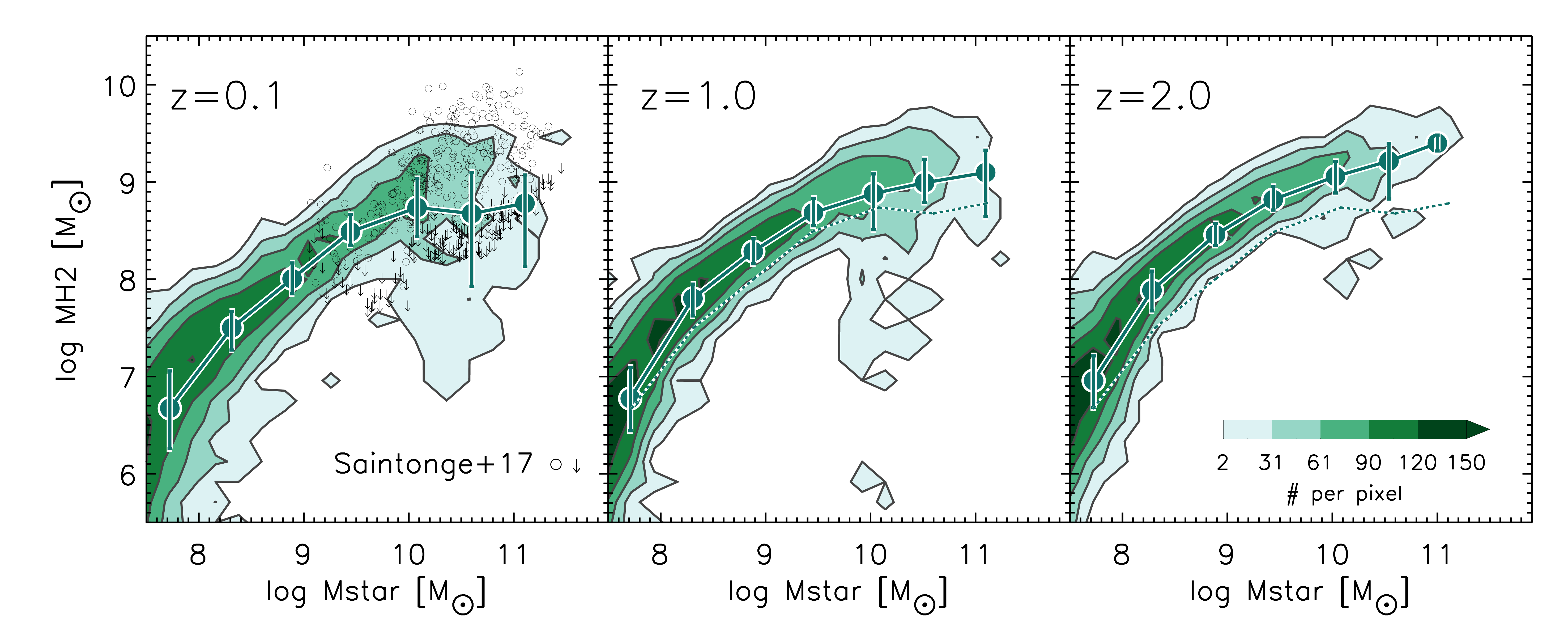}
\caption
{Molecular hydrogen mass as a function of stellar mass for galaxies in our simulation.
In addition to contours, solid circles display the median $\rm{MH2}$ within bins of stellar mass, along with the 25\%-75\% percentiles for model galaxies with $\rm{MH2} > 0$. To facilitate comparison, the dotted line in the central and right panels corresponds to the $z=0$ relation. The colour contours delineate grids containing a specific number of galaxies as indicated in the colour bar, utilizing 0.2 dex-wide bins along both the x and y axes. Empty circles in the left panel represent the xCOLD GASS sample reported in \citep{Saintonge2017}; non-detection in this sample is represented by downward arrows (0.01 < redshift < 0.05).}
\label{fig:h2msm}
\end{figure*}

In Fig. \ref{fig:h2msm}, we depict the relationship between the stellar mass and the \hmol mass content of galaxies at $z=0$, 1, and 2. We also plot the $z=0$ relationship in the central and right panels to facilitate comparison across these redshifts.  It can be observed that galaxies of equal mass exhibit higher \hmol content as redshift increases from $z=0$ to $z=2$, particularly at higher stellar masses (a factor $\sim$ 4).
At redshift $z=0.1$, we compare our predicted \hmol masses with the determinations by \mbox{\cite{Saintonge2017}}. In their study, galaxies without CO detections are assigned upper limits, represented by downward arrows in the figure.
The contours representing our full population of galaxies align well with the observational data. The median \hmol mass per stellar mass bin falls below observational detections at the highest stellar masses. However, considering the non-detections, our median might be a fair representation of the real population, within which galaxies undergoing quenching are more likely to go undetected.

\begin{figure*}
\centering
\includegraphics[width=\textwidth]{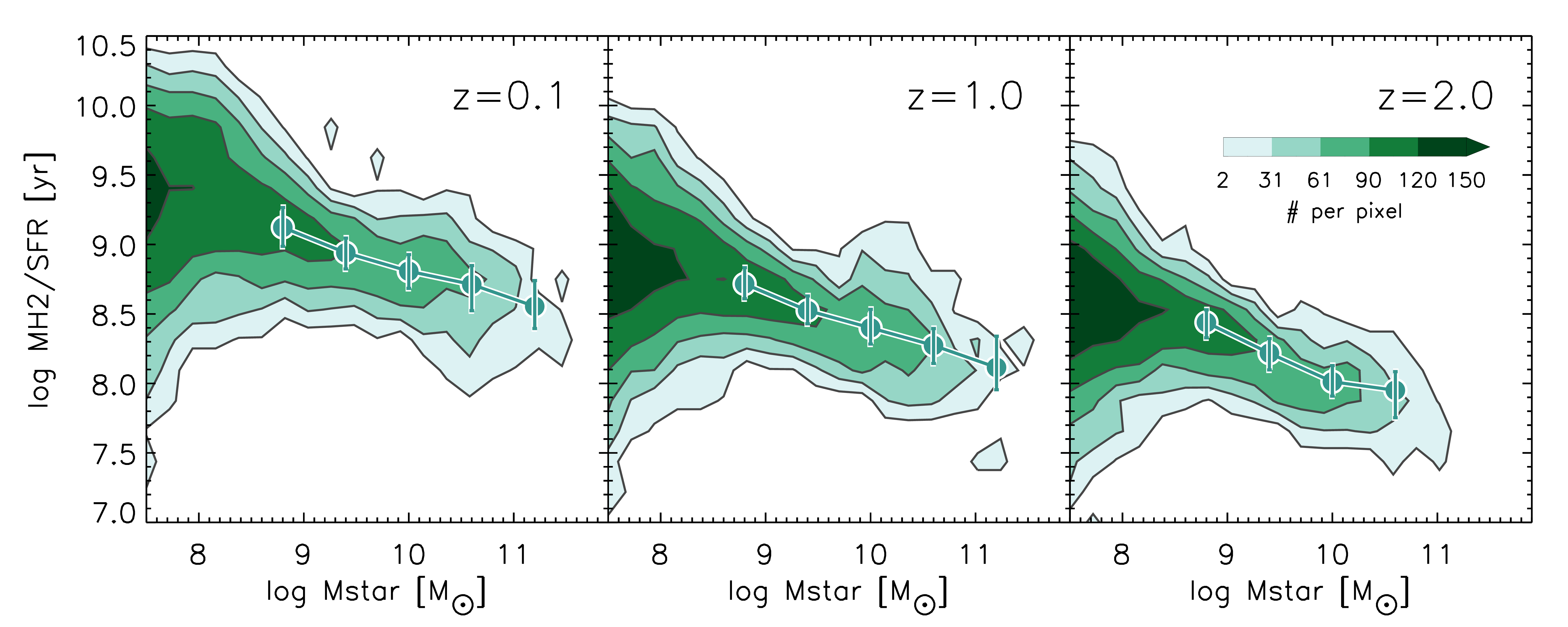}
\caption
{\revv{Depletion time as a function of stellar mass for model galaxies.
In addition to contours, solid circles display the median depletion time within bins of stellar mass, along with the 25\%-75\% percentiles for model galaxies with $\rm{MH2} > 0$. The colour contours delineate grids containing a specific number of galaxies as indicated in the colour bar, utilizing 0.2 dex-wide bins along both the x and y axes.}}
\label{fig:dept}
\end{figure*}

Finally, we notice that the MH2 vs Mstar relationship bends in the log-log plane at high stellar masses,  driving the bending of the MS  discussed in \ref{sec:MS} (see Fig. \ref{fig:mainseq}). \rev{The latter bending is somewhat less pronounced than the former. Indeed, the depletion time MH2/SFR tends to decrease with increasing mass in our simulations}, \revv{as shown in Figure \ref{fig:dept}}. \rev{This occurs because more massive galaxies feature, on average, a shorter free fall time $t_{\rm ff}$ in the cold phase of MP particles (see Eq. \ref{eq:SFlaw}).
These galaxies, which tend to be spheroid-dominated, are quenched because AGN feedback decreases their SF fuel and not because morphological quenching \citep[e.g.][]{martig2009,gensior2020} stabilises the cold gas against gravitational collapse.
} 

\begin{figure*}
\centering
\includegraphics[width=\textwidth]{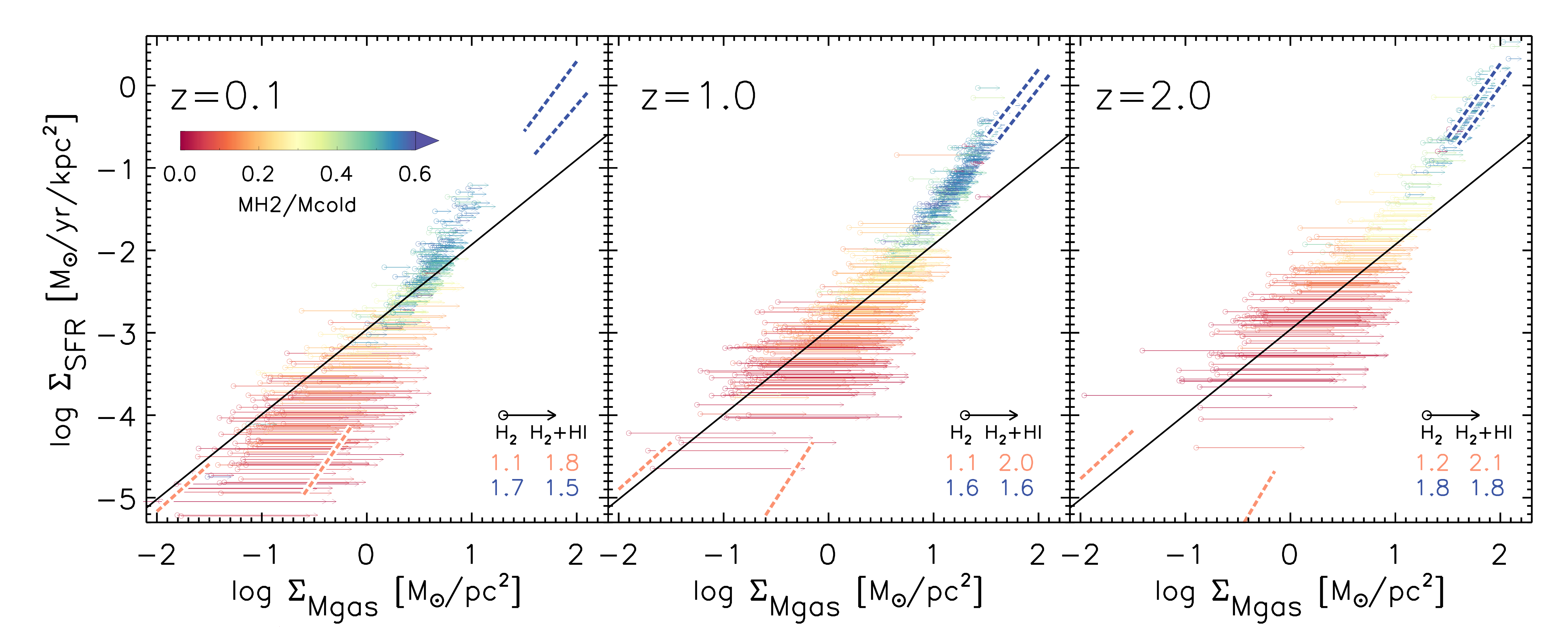}
\caption[width=\textwidth]
{Integrated KS relation. The tail of the arrows represents $\Sigma_{\rm H2}$, while the head indicates $\Sigma_{\rm H2+HI}$. We show the results for galaxies with stellar masses Mstar > $10^8 \msun$ and bulge-to-total ratios  B/T < 0.5. The colour bar encodes the \hmol fraction of each galaxy. Red / blue dashed line segments are linear fits in the log-log plane for galaxies with \hmol fraction lower / greater than 0.5. The corresponding slopes are labelled in each panel for \hmol and ${\rm H_2+HI}$ gas phases.
Solid line represents one of the fits (slope n=1.03) from \cite{delosreyes2019} to the $\Sigma_{\rm H2}$-$\Sigma_{\rm SFR}$ relation obtained from a compilation of observational samples of local spirals.  
}
\label{fig:kenh2}
\end{figure*}

\begin{figure*}
\centering
\includegraphics[width=\textwidth]{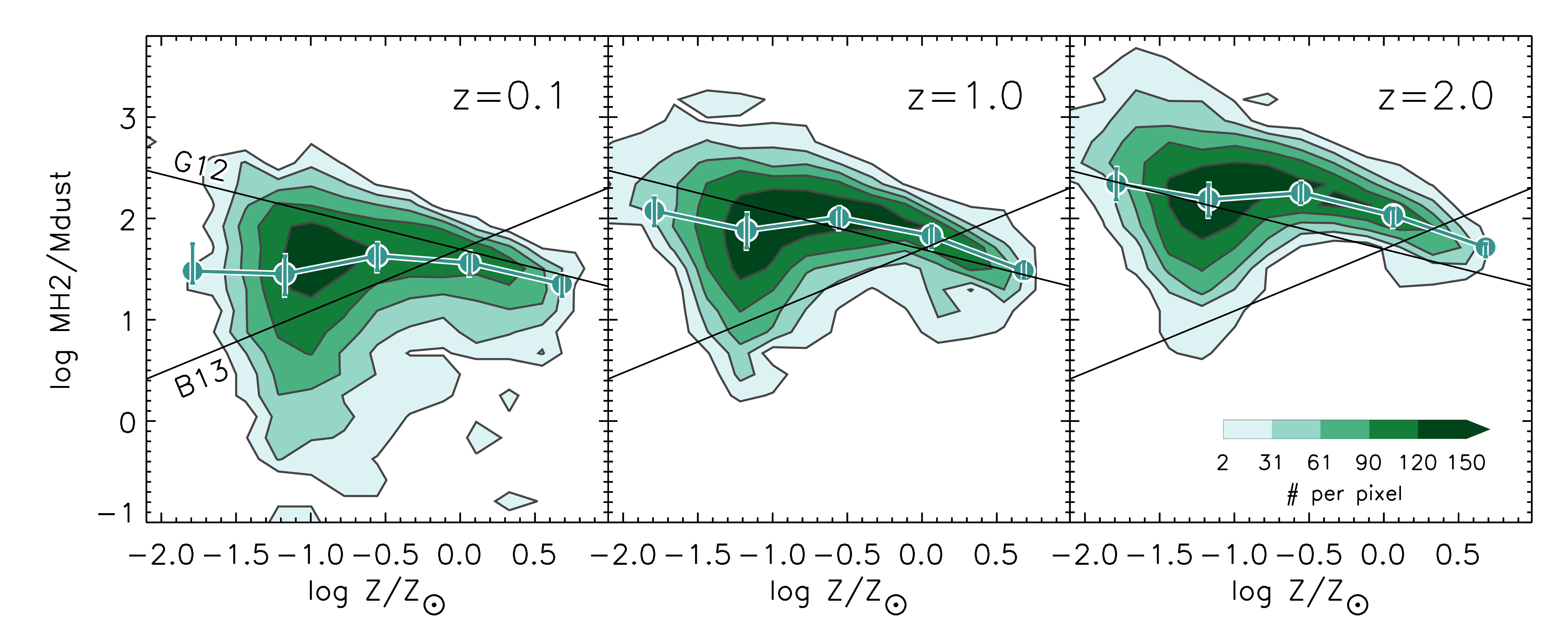}
\caption[width=\textwidth]
{Contours depict the $\rm{H_2}$-to-dust ratio as a function of gas metallicity for all model galaxies. MS galaxies ($\rm{\mid log(SFR)-log(SFR_{MS}) \mid < 0.5}$) at the given redshift are represented by the circles, which depict the median $\rm{H_2}$-to-dust ratio within stellar mass bins, accompanied by the 25\%-75\% percentiles. The two lines labelled as G12 and B13 correspond to fits obtained in \cite{bertemes18} when using the CO-to-$\rm{H_2}$ conversion factors derived in \cite{genzel12} or \cite{bolatto13}, respectively. The colour contours enclose grids containing a specific number of galaxies as indicated in the colour bar, utilizing 0.2 dex-wide bins along both the x and y axes. 
}
\label{fig:h2tdZ}
\end{figure*}


\begin{figure*}
\centering
\includegraphics[width=\textwidth]{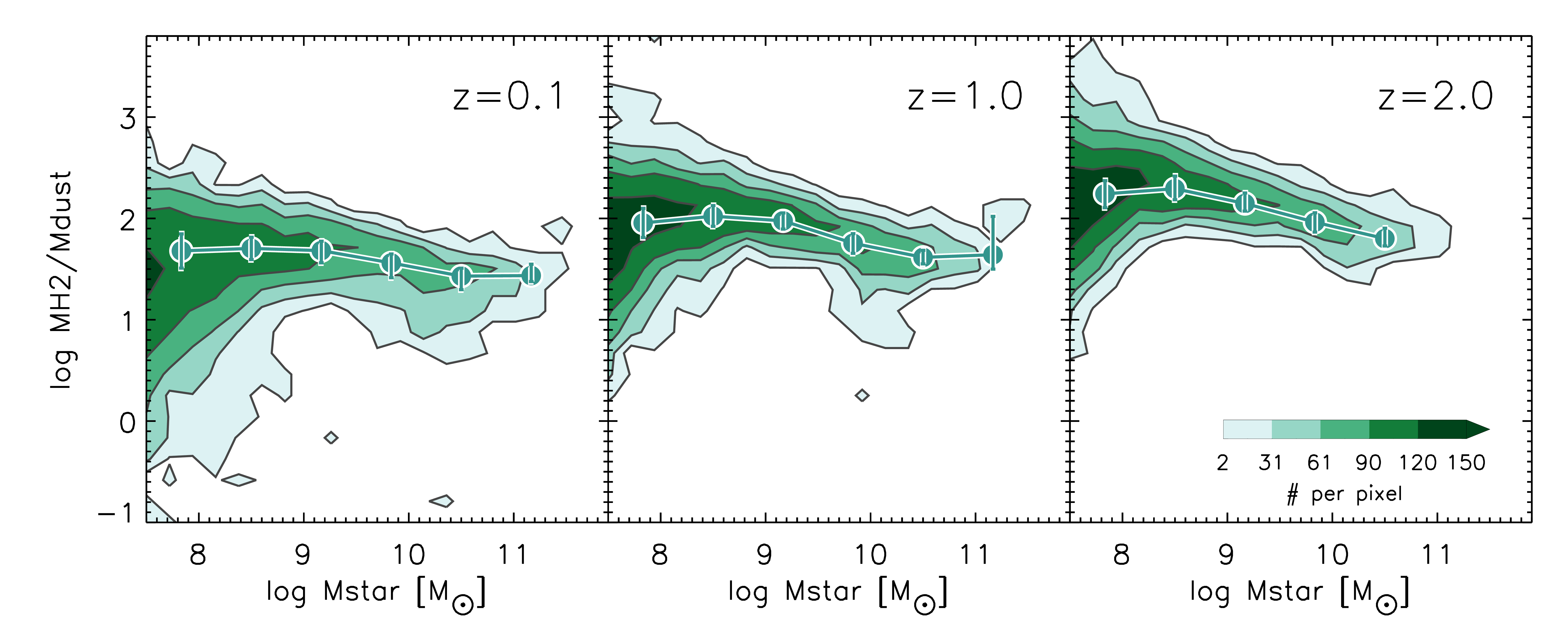}
\caption[width=\textwidth]
{Contours depict the $\rm{H_2}$-to-dust ratio as a function of stellar mass for all model galaxies. MS galaxies ($\rm{\mid log(SFR)-log(SFR_{MS}) \mid < 0.5}$) at the given redshift are represented by the circles, which depict the median $\rm{H_2}$-to-dust ratio within stellar mass bins, accompanied by the 25\%-75\% percentiles. The colour contours enclose grids containing a specific number of galaxies as indicated in the colour bar, utilizing 0.2 dex-wide bins along both the x and y axes. 
}
\label{fig:h2tdsm}
\end{figure*}

\begin{figure*}
\centering
\includegraphics[width=\textwidth]{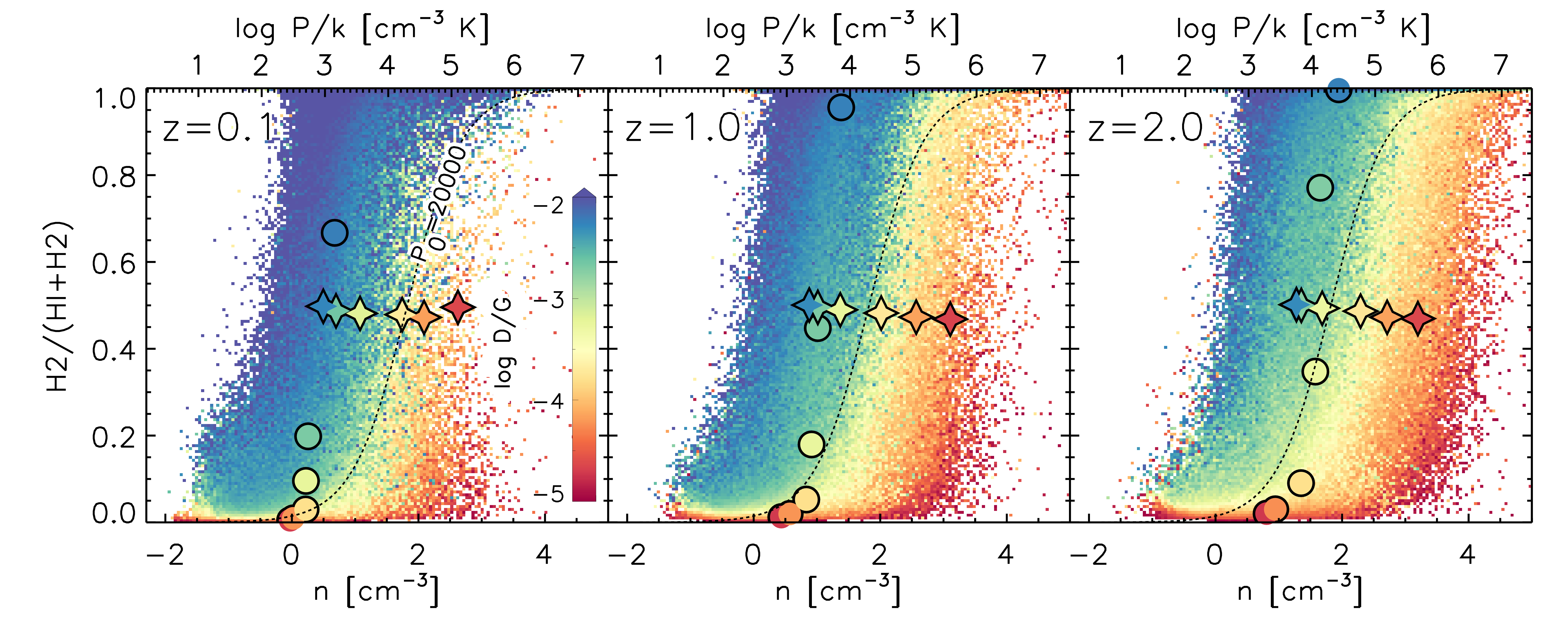}
\includegraphics[width=\textwidth]{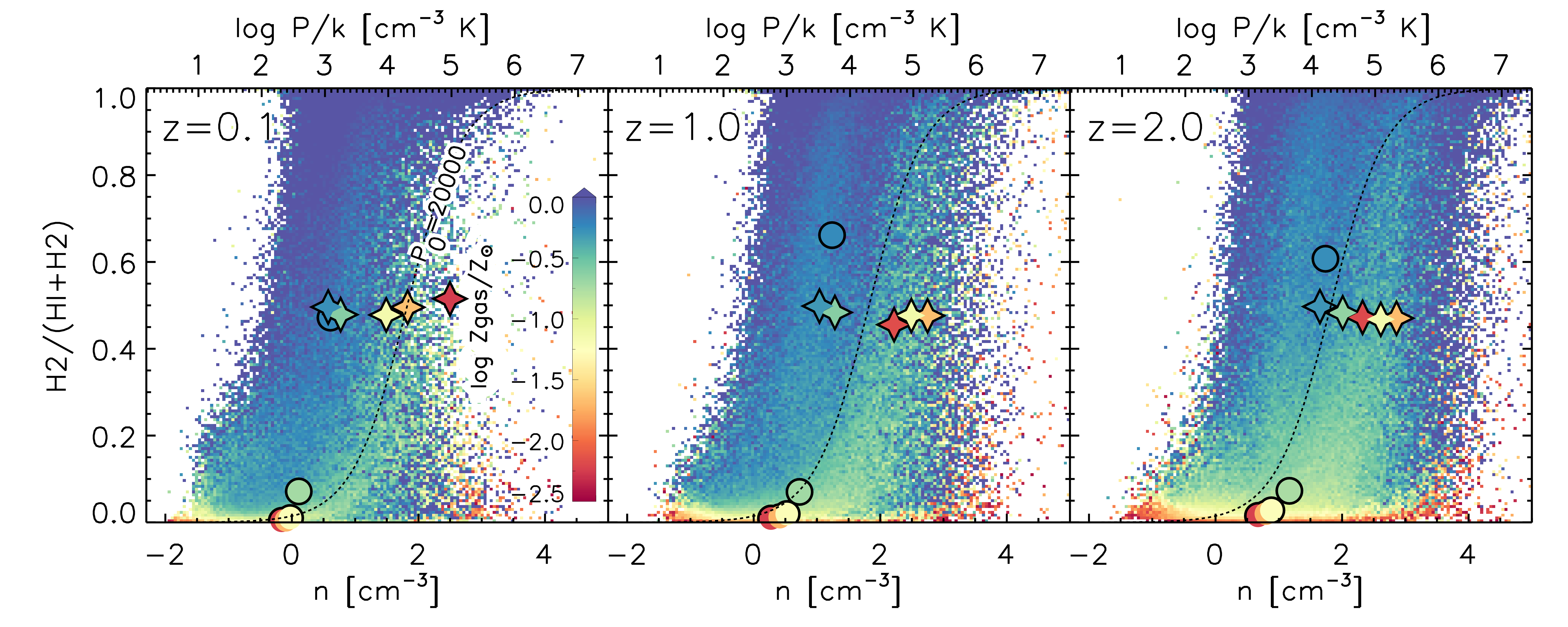}
\caption[width=\textwidth]
{Transition from atomic to molecular hydrogen in MP particles illustrated for the three analysed redshifts as a function of the sub-grid number density of the cold phase. The latter is assumed to have a temperature 300K hence the pressure is proportional to the number density and shown in the top axis (k is the Boltzmann constant)). The colour code shows the mean D/G mass ratio (top panel) and the gas phase metallicity (metals-to-gas mass ratio, bottom panel) of the gas particles falling within each pixel (150 pixels per axis). colour-coded circles represent the median \hmol fraction and number density for particles within 0.5 dex bins of log(D/G) (top panel) and $\rm{log(Zgas/Z\sun)}$ (bottom panel). Meanwhile, stars indicate the same, but exclusively for particles in the transition region with \hmol fraction between 0.4 and 0.6.}
\label{fig:moltrans}
\end{figure*}


\section{The Kennicutt-Schmidt relation}
\label{sec:KS}
\rev{Over the decades, a well-defined correlation between the disc-averaged gas and SFR surface densities in galaxies has been identified and extensively studied \citep{schmidt59,kennicutt1989}. This relationship is now commonly referred to as the (in its disc-averaged form, either global or integrated) Kennicutt-Schmidt (KS) law. This law holds over several orders of magnitude and has inspired most of the sub-resolution prescriptions adopted in galaxy formation models, including ours. More recently, observational studies have concentrated on analysing this correlation at resolved scales and distinguishing between the atomic and molecular gas surface densities. The general result is that the KS relation holds even within galaxies, and it is primarily driven by the molecular component \citep[see][and references therein]{delosreyes2019}. The latter point confirms that molecular clouds are closely linked to the fuel for star formation.}

\rev{The integrated KS relation for our model galaxies at various redshifts is shown in Fig. \ref{fig:kenh2}. For each galaxy, an arrow indicates the shift from its position in the plot when considering only molecular gas (the molecular Kennicutt-Schmidt, mKS) to its position when both atomic and molecular gas are included (the total Kennicutt-Schmidt, KS). The arrow colour encodes the \hmol mass fraction of the galaxy. Both the colour and the length of the arrows indicate a clear increase of this fraction with the surface density. Moreover, at low densities, the arrowheads display a somewhat larger horizontal dispersion than the tails: the molecular gas alone is more closely related to SFR than the total gas. Conversely, at high surface density, HI is sub-dominant, and the total mKS approaches the total KS relation. 
The black solid line corresponds to a fit to Local Universe data compiled by \cite{delosreyes2019}, consolidating information from 114 publications on the KS.  Its slope is $\rm n=1.03$ and is reported in all panels for reference. To approximate the observational sample, the simulated sample includes galaxies with stellar masses greater than $\rm{10^8\msun}$ and bulge-to-total ratios $\rm{B/T < 0.5}$. We apply the same selection criteria for $z=1$ and $z=2$ galaxies, shown in the central and right panels.}

The simulation captures the integrated mKS relation \rev{(the arrow tails)} at redshift $z=0$, exhibiting a reasonable alignment with the $\rm n=1$ slope at $\Sigma_{\rm H2} \lesssim 4 \, \rm{M}_\odot \,\rm{pc}^{-2}$. 
Above this limit, a steepening of the relationship is evident. At higher redshift $z=1$ and 2 (central and right panels of Fig. \ref{fig:kenh2}), we find that the two-slope shape of the relation is preserved, but the normalization increases by a factor $\sim 2$ and 4, respectively.

\rev{Recently, \cite{teng24} analysed the central  1.5 kpc regions of 80 galaxies from the PHANGS-ALMA survey \citep{Leroy2021}, by adopting a CO-to-$\rm{H_2}$ conversion factor depending on the local CO velocity dispersion at a 150 pc scale. 
They concluded that at the high $\Sigma_{\rm H2}$ levels reached in galaxy centres, the mKS relation becomes steeper, a phenomenon that goes undetected when a fixed CO-to-$\rm{H_2}$ conversion factor is used.
While our estimate of the mKS is integrated over the entire galaxy, it is suggestive that we get a clear steepening to a similar slope at the high $\Sigma_{\rm H2}$ end.}

In our model, the steepening of the mKS relation at higher $\Sigma_{\rm H2}$ has the following interpretation. The SFR is computed in MP particles using Eq. \ref{eq:SFlaw}. Thus, we expect that $\Sigma_{\rm SFR} \propto  \Sigma_{\rm H2}/t_{\rm ff}$, where the \rev{free-fall} time is computed using the density of the cold phase $t_{\rm ff} \propto \rho_{\rm c}^{-0.5}$. The molecular fraction tends to be low at low $\Sigma_{\rm H2}$ values, as illustrated by the colour coding in the figure. Then, the \rev{free-fall} time is substantially unaffected by the \hmol density, being the latter a sub-dominant component of the cold gas (see scheme in Fig. \ref{fig:MPparticle}). Conversely, as $\Sigma_{\rm H2}$ increases, so does the \hmol fraction, and the \rev{free-fall} time tends to scale progressively as $t_{\rm ff} \propto \rho_{\rm H2}^{-0.5}$. In the former regime, we then expect a linear proportionality between the \hmol and the SFR densities, steepening towards $\Sigma_{\rm SFR} \propto  \Sigma_{\rm H2}^{1.5}$ in the latter regime. 
Moreover, the correlation shows a larger dispersion in the former regime. When the \hmol fraction is low, the density and the free-fall time are determined mainly by HI. As a result, in this regime $\Sigma_{\rm SFR}$ depends on the density of two species, HI and $\rm{H2}$, with the former exhibiting dispersion at any given value of the latter. Conversely, at high $\Sigma_{\rm H2}$, where almost all the cold mass is made of $\rm{H_2}$, SFR computation is based solely on one mass species, reducing its dispersion.

\section{\hmol mass determination from dust}
\label{sec:h2dust}
As mentioned before, the emission of \hmol molecules is not directly measured. Therefore, any evaluation of the molecular gas content in galaxies is limited to indirect methods. As molecules predominantly form on the surface of grains, which also act as a shield against dissociating radiation, one such method involves utilizing dust as a tracer for molecular gas. 

In a recent work, by using star-forming galaxies with stellar masses $>10^{10} \msun$ and $z\lesssim 0.2$, \cite{bertemes18}  establish a formula to derive molecular masses from dust masses (their eq. 15): $\rm{log(M_{mol}) = log(M_{dust}) + 1.83 + 0.12[(12+log(O/H))+8.67]}$.  
This involved computing CO-based molecular masses, which introduces a dependency on metallicity through the conversion factor CO-to-$\rm{H_2}$. 
In their work, the coefficient of 0.12 multiplying the metallicity term is obtained by adopting a geometric mean of the CO-to-$\rm{H_2}$ conversion factors derived in \cite{genzel12} (G12) and \cite{bolatto13} (B13). This coefficient changes to -0.37 or 0.61 if the geometric mean is replaced by the CO-to-$\rm{H_2}$ conversion factors derived in G12  and B13, respectively. Their formulas yield similar values around the solar metallicity but diverge significantly at lower and higher Z. In G12, the CO-to-$\rm{H_2}$ conversion factor is derived using a mixed sample of main-sequence star-forming galaxies with $z\gtrsim 1$ plus five Local Group galaxies. The whole sample metallicity range is  $ \rm{-0.9 ~ \lesssim ~ log(Z/Z_{\odot}) \lesssim 0.3}$. On the other hand, B13 formula is derived from theoretical grounds.

In Figure \ref{fig:h2tdZ}, contours show $\rm{H_2}$-to-dust ratio as a function of metallicity for the simulated galaxies. We over-plot the two lines corresponding to the \cite{bertemes18} $\rm{H_2}$-to-dust equation (without He fraction) but using the mentioned G12 and B13 conversion factors.
The circles connected by solid lines illustrate the $\rm{H_2}$-to-dust ratios plotted against metallicity for galaxies within the MS, specifically those with a deviation $\rm{\mid log(SFR)-log(SFR_{MS}) \mid < 0.5}$. The MS at each redshift were determined by fitting a functional form as the one proposed in \cite{popesso23}, which is suitable for our model (see Subsection \ref{sec:MS} and Fig. \ref{fig:mainseq}).
The general trend across all redshifts suggests that galaxies with higher metallicities exhibit smaller $\rm{H_2}$-to-dust ratios, with a higher normalization associated with higher redshifts. This trend is at odd with Eq. 15 in B18, based on the geometric mean of G12 and B13 prescription and featuring a positive slope. However, it aligns with their findings when only the G12 prescription is considered. The match is particularly good at $z=0$ and $z=1$, where the curve for model galaxies approaches G12 line in the range of metallicities used by that study. 

The relationship between the $\rm{H_2}$-to-dust ratio and stellar mass is illustrated in Fig. \ref{fig:h2tdsm}. As in the previous figure, the $\rm{H_2}$-to-dust fraction for the global population is smaller now than in the past. In our simulated galaxy sample, the most massive galaxies ($10.75 \lesssim \rm{log(Mstar/M_\odot)} \lesssim 11.25$) exhibit an $\rm{H_2}$-to-dust ratio of approximately 65 at $z=2$ and 30 at $z=0$. While at the lowest mass end, the typical $\rm{H_2}$-to-dust ratio is about 160 at $z=2$ and 50 at $z=0$. However, the dispersion in the $\rm{H_2}$-to-dust values at these lower masses is significantly higher compared to the high mass end. Consistently, this feature is also present at low metallicities in Fig. \ref{fig:h2tdZ} .


\section{The atomic to molecular transition}
\label{sec:moltrans}
Fig. \ref{fig:moltrans} shows the dependence of MP gas particle \hmol fraction on the cold component sub-grid number density at redshift 0.1,1, and 2. Since the cold gas in MP is assumed to have a given temperature of 300 K, its number density translates univocally into pressure, as shown on the top axis scale. The dotted black line is the relationship used in previous MUPPI works to estimate the molecular fraction from the gas pressure, inspired by the \cite{Blitz2006} findings.\footnote{\rev{We recall that in the computation of \hmol formation rate (Eq. \ref{eq:hmolform}), we assume a clumping factor $C_p=10$,  effectively amplifying the average density of the cold phase, as shown on the plot axis, by a factor of $\sqrt 10$. This is intended to approximate the inhomogeneity of the medium.}}
The colour coding in the top panel corresponds to the D/G ratio of particles. The fraction of \hmol increases with particle density and D/G ratio, as can be understood based on the linear dependence of the \hmol formation rate on atomic hydrogen and grain densities (see Eq. \ref{eq:RG2}). To illustrate this trend, the median \hmol fraction and number density are presented for particles grouped into six bins of D/G ratios (each spanning 0.5 dex), represented by coloured coded circles. These circles show that the typical \hmol fraction decreases as redshift decreases for particles within a specific D/G bin. 
For instance, the \hmol fraction of the most "dusty" particles in our simulations, characterized by $\rm{-2.5<log(D/G)}$, exhibit a median value of approximately 1, 0.95, and 0.7 at redshifts $z=2$, 1, and 0, respectively.

The transition from low to high molecular fractions at a given value of D/G is sharp and tends to shift to slightly higher values of $\rm n$ with increasing redshift. The latter is evidenced by the coloured stars, which show the median number density for particles of a given D/G having \hmol fraction between 0.45 and 0.55. This effect can be attributed to the decreasing importance of small grains in the dust budget at high redshift. While large grains dominate the dust mass, small grains dominate the surface area, and the \hmol formation process is a surface process. By taking into account these considerations, a practical application could involve developing prescriptions to estimate the typical \hmol fraction within a specific region of the ISM for models where \hmol evolution is not explicitly computed. This estimation would consider factors such as gas density and the budgets of small and large grains.

Finally, in the bottom row of Fig. \ref{fig:moltrans}, the pixels are colour-coded according to gas metallicity. It is evident that gas metallicity exhibits a much weaker correlation with the molecular fraction at a given density than D/G (top panel).
This weak correlation is not unexpected, given that the relationship between dust and metal content is not straightforward: \rev{Metal abundance does not directly control the formation rate of $\rm{H_2}$. Instead, it is primarily governed by dust abundance and gas density.}
Additionally, when considering the content of small grains, which, as mentioned earlier, significantly impact the rate of \hmol formation, the complexity of this relationship becomes even more apparent. Nevertheless, it can still be envisaged that, as expected, the molecular transition occurs typically at higher densities when the metallicity decreases.




\section{Summary and conclusions}
\label{sec:conclusions}

From the perspective of galaxy formation models, one of the most necessary tasks for the coming years is integrating cosmological simulations with models capable of reproducing the creation and evolution of the gaseous molecular phase. In this work, we have taken a significant step towards this direction within the MUPPI sub-resolution model \citep[see][and references therein]{valentini23} to describe star formation from a multi-phase ISM, taking advantage of our recent incorporation of dust evolution in cosmological simulations \citep{granato21}.
By establishing a connection between SF, ISM metal enrichment, dust production, and dust-promoted \hmol formation, we effectively model the
star formation-dust-\hmol loop present in nature,
resulting in a more physically grounded model for non-primordial ISM.
\\

\noindent 
The main findings of this study can be summarized as follows:
\begin{itemize}

    \item Our model, which is calibrated based on the cosmic star formation rate density, the stellar mass function at $z=0$, and the stellar mass-halo mass relation at $z=0$, satisfactorily matches several other observational data. This includes the stellar mass function at $z=1$ and $z=2$, the dust and \hmol mass functions at $z=0$, $z=1$, and $z=2$, the cosmic evolution of \hmol density, and the relation between molecular and stellar mass in galaxies (see Figures \ref{fig:sfrd} to \ref{fig:h2msm}).

    \item The integrated molecular Kennicutt-Schmidt relation emerges in our model galaxies as early as redshift $z\sim 2$, although with a higher normalization than at redshift 0 (Fig. \ref{fig:kenh2}). The slope remains close to unity across the considered redshifts for surface densities below $\rm{\sim 5\ \msun/pc^2}$. However, at higher surface densities, we observe a slope more consistent with 1.5,
    \rev{which is reminiscent of a recent observational finding concerning the dense central 1.5 kpc regions of galaxies \citep{teng24}}. The normalization of the simulated Kennicutt-Schmidt relation at redshift zero aligns well with determinations from the Local Universe. The emergence of a molecular KS relation and its steepening at high molecular density in our simulations are natural consequences of the adopted SF law (Equation \ref{eq:SFlaw}).  However, the normalization of the curve results from the calibration process, and its evolution can be considered a model prediction.
    
    \item The $\rm{H_2}$-to-dust ratio diminishes with increasing metallicity across all examined redshifts (Fig \ref{fig:h2tdZ}).  Our model favours the G12 CO-to-$\rm{H_2}$ conversion factor. A decrease of the $\rm{H_2}$-to-dust ratio is also observed with increasing stellar masses (Fig \ref{fig:h2tdsm}), particularly at the highest masses. The normalization of these curves decreases from $z=2$ to $z=0$, implying that the $\rm{H_2}$-to-dust fraction for the global population is now smaller than it was in the past.
    Galaxies in our most massive bin ($10.75 \lesssim \rm{log(Mstar/M_\odot)} \lesssim 11.25$) have an $\rm{H_2}$-to-dust ratio of $\sim 65,\ 30$ at $z=2,\ 0$ respectively. At the lowest mass end, the typical $\rm{H_2}$-to-dust varies from $\sim 160$ at $z=2$ to $\sim 50$ at $z=0$, though the dispersion in the $\rm{H_2}$-to-dust values at these masses (or metallicities) is much higher than at the high mass end (or metallicities).

    \item On a particle-by-particle basis, the transition from atomic to molecular hydrogen at a specific D/G value is abrupt (Fig, \ref{fig:moltrans}). It tends to shift towards higher number densities with increasing redshift. This behaviour is primarily influenced by the crucial role of small grains, which dominate the surface per unit mass and whose relative importance diminishes with increasing redshift compared to large grains. 
    
    \item The typical \hmol fraction decreases as redshift decreases for particles with a given D/G (Fig, \ref{fig:moltrans}).
\end{itemize}

In upcoming studies, we plan to devise methods for estimating the typical \hmol fraction within specific regions of the ISM to be incorporated in models that do not explicitly compute \hmol or dust evolution. To comprehensively describe galaxy evolution, especially at high redshifts, we also aim to model the direct formation of \hmol in the gas phase, a process relevant in dust-free conditions. 
   
\begin{acknowledgements}
We thank the anonymous referee for carefully reading our manuscript and the interesting comments provided.
This project has received funding from the Consejo Nacional de Investigaciones Cient\'ificas y T\'ecnicas (CONICET) (PIP-2021-11220200102832CO), from the Agencia Nacional de Promoción de la Investigación, el Desarrollo Tecnológico y la Innovación de la Rep\'ublica Argentina (PICT-2020-03690), and from the European Union's HORIZON-MSCA-2021-SE-01 Research and Innovation Programme under the Marie Sklodowska-Curie grant agreement number 101086388 - Project (LACEGAL). MV is supported by the Italian Research Center on High Performance Computing, Big Data and Quantum Computing (ICSC), project funded by European Union - NextGenerationEU - and National Recovery and Resilience Plan (NRRP) - Mission 4 Component 2, within the activities of Spoke 3, Astrophysics and Cosmos Observations, and by the INFN Indark Grant.
Simulations have been carried out at the computing centre of INAF (Italy). We acknowledge the computing centre of INAF-Osservatorio Astronomico di Trieste, under the coordination of the CHIPP project \citep{bertocco2019,Taffoni2020}, for the availability of computing resources and support.     
\end{acknowledgements}

%
   \bibliographystyle{aa_url} 
   \bibliography{bib, bibmilena} 
%

\begin{appendix}

\section{Refinements to MUPPI}
\label{app:refMUPPI}
This section illustrates the effect of the two most relevant modifications made to MUPPI to achieve a cosmological simulation box exhibiting galaxy statistical properties consistent with observational data, as discussed in Section \ref{sec:MUPPImodifications}.

Wind particles are designed to sample galactic outflows. The details about their behaviour can be found in \cite{murante15} and \cite{ valentini17}.  
A gas particle exits its multi-phase stage after a maximum allowed time given by the \rev{free-fall} time of the cold gas. 
\rev{When a gas particle exits a multi-phase stage, it has a probability $P_{\rm kin}$ (a model parameter) of being kicked and becoming a wind particle for a time interval $t_{\rm wind}$. 
Wind particles are decoupled from hydrodynamic forces. While they are affected by radiative cooling, they cannot contribute to the artificial thermal conduction implemented in the P-Gadget SPH. Moreover, they receive kinetic energy from neighbouring star-forming gas particles, which increases} their velocity along their least resistance path since they are kicked against their own density gradient. 
The wind stage can conclude before $t_{\rm wind}$ whenever the particle density drops below a chosen density threshold. 

\begin{figure}
\centering
\includegraphics[width=0.48\textwidth]{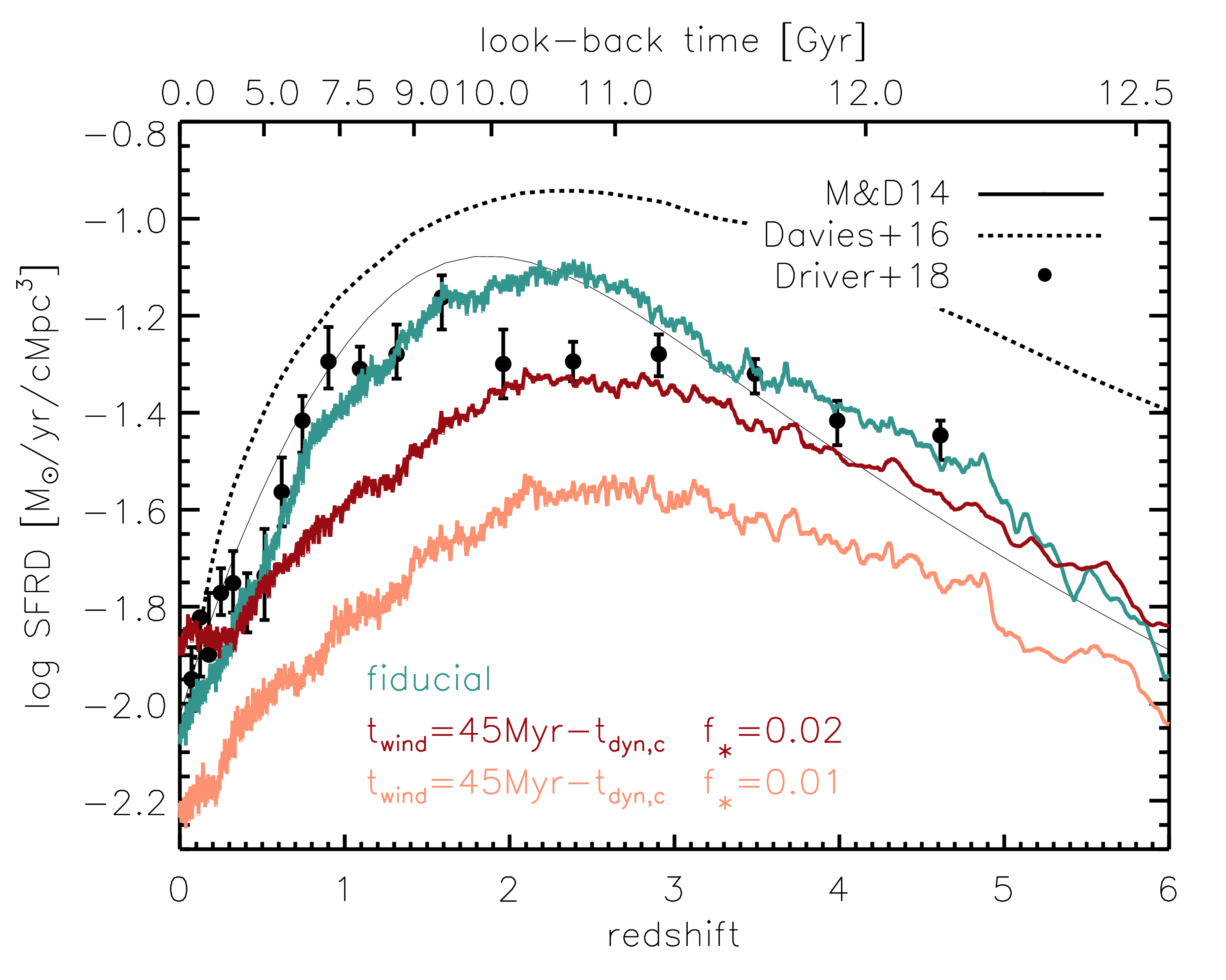}
\includegraphics[width=0.48\textwidth]{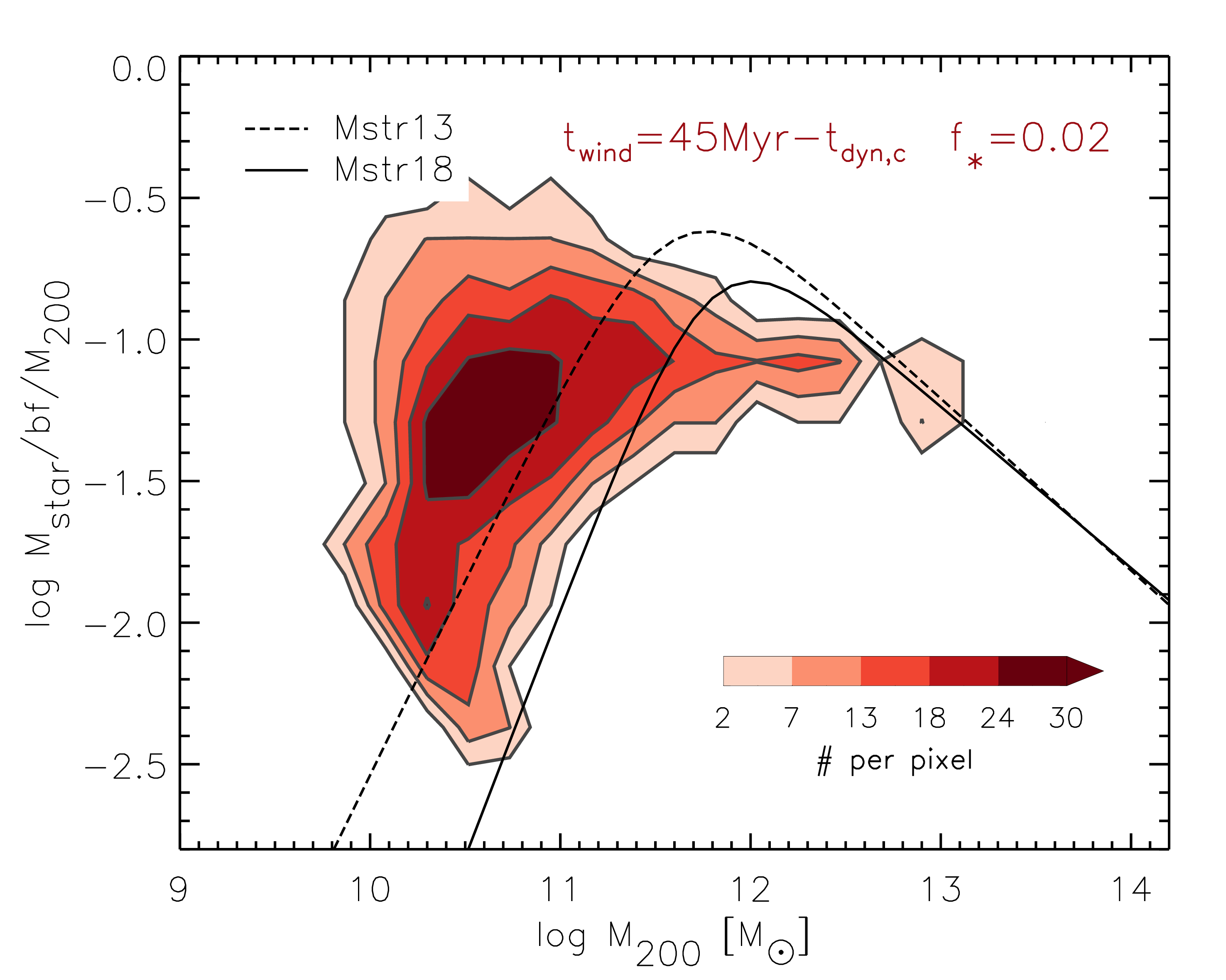}
\includegraphics[width=0.48\textwidth]{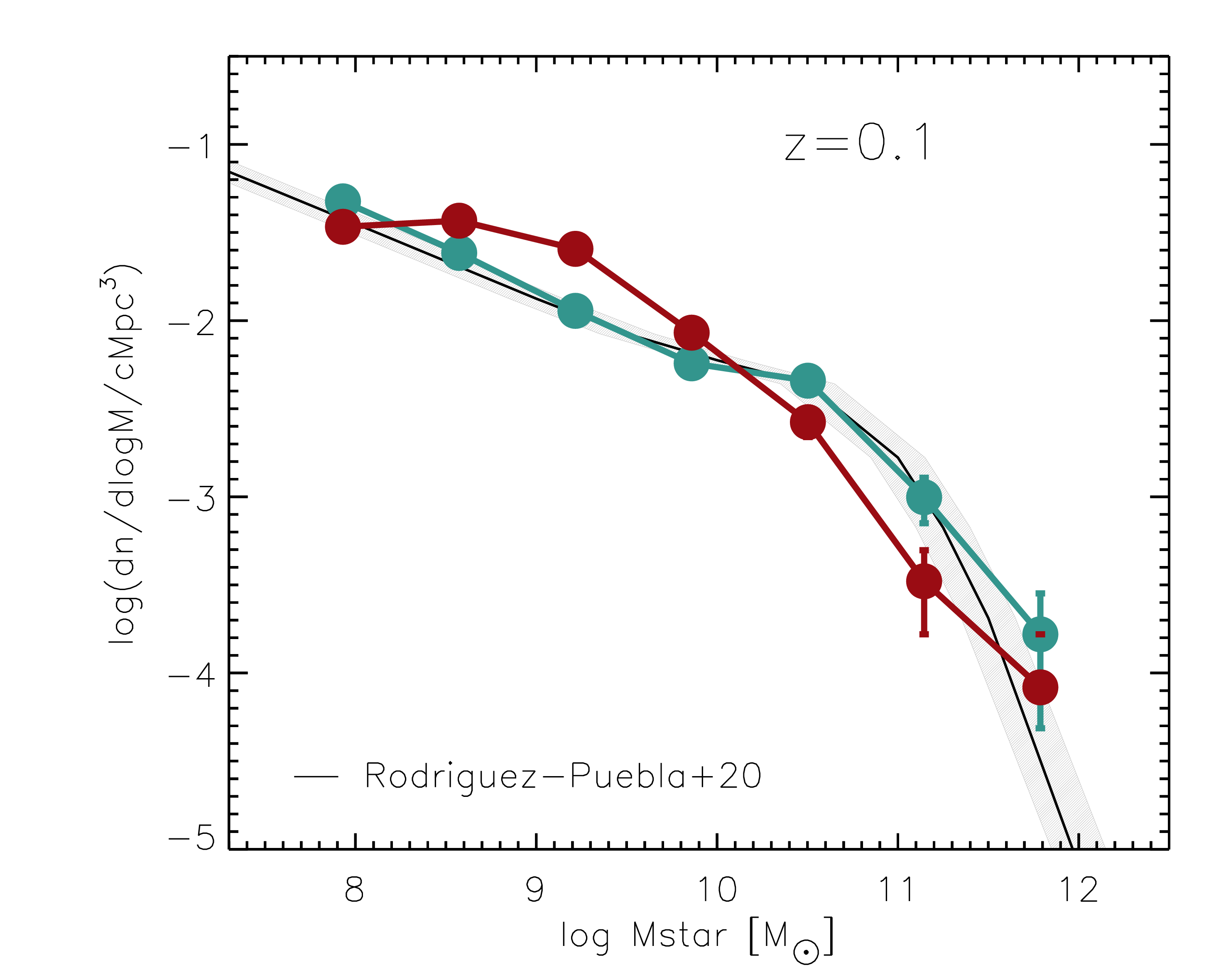}
\caption[width=\textwidth]
{Cosmic SFRD from our reference simulation is compared with two alternative runs for one of the simulated boxes in the top panel. Both these runs do not scale the particle wind phase duration with $\sigma_{DM}$, as in the fiducial model, instead adopting $\rm{t_{wind}=45Myr-t_{ff,c}}$. They differ for the adopted SF efficiency $f_*$, which is either 0.01 or 0.02. Observational determinations are the same as in Fig. \ref{fig:sfrd}. In the middle and bottom panels we show the SHMR and the SMF for the $f_*$=0.02 case.}
\label{fig:oldsetup1}
\end{figure}

Figure \ref{fig:oldsetup1} shows the importance of scaling a particle duration in the wind phase $t_{\rm wind}$ with the velocity dispersion of its neighbouring DM particles, \rev{as in Equation \ref{eq:twind}.} 
In the top panel, we present the SFRD for our reference model alongside two alternative runs for comparison. The latter two do not scale $t_{\rm wind}$ with $\sigma_{DM}$, instead adopting $t_{\rm wind} = 45\ \text{Myr} - t_{\rm ff, c}$ as in previous MUPPI runs, and in particular as in \cite{parente22}. They differ in the adopted SF efficiency $f_*$. In the run with $f_*$=0.01, the efficiency is the same as in the fiducial version of the model. However, the very low SFRD results in this solution being completely unsatisfactory. Therefore, we also show a simulation with $f_*$ increased to 0.02, as in previous versions of MUPPI. The SFRD for this latter case approaches the observed SFRD, still without satisfactorily reproducing it. Further increasing $f_*$ is not viable because, as evident in the middle and bottom panels, this assumption for computing $t_{\rm wind}$ results in a misrepresentation of how halos are populated by galaxies and an inaccurate count of galaxies of a given mass, issues that are not solved by a further increase in $f_*$.

\begin{figure}
\centering
\includegraphics[width=0.48\textwidth]{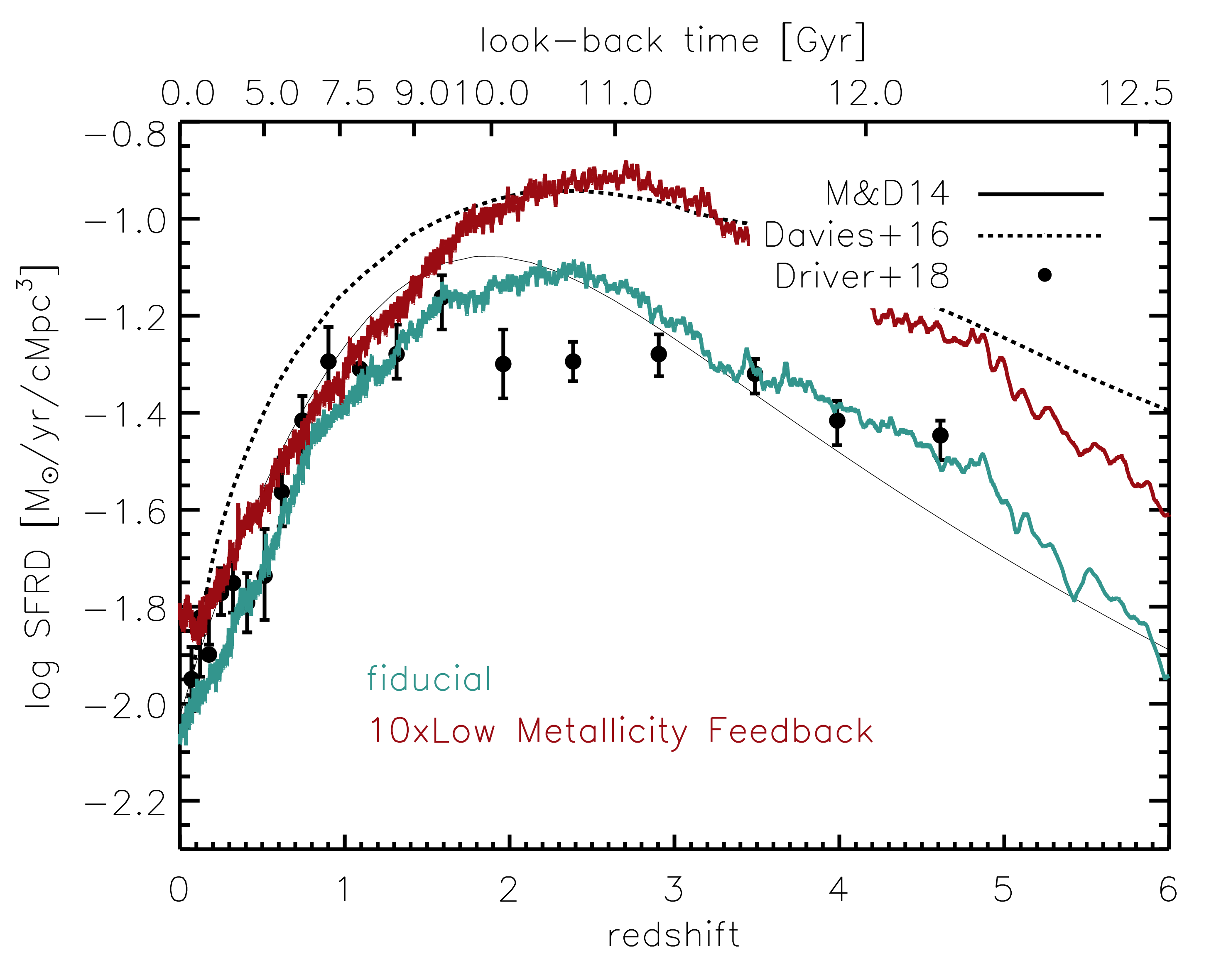}
\includegraphics[width=0.48\textwidth]{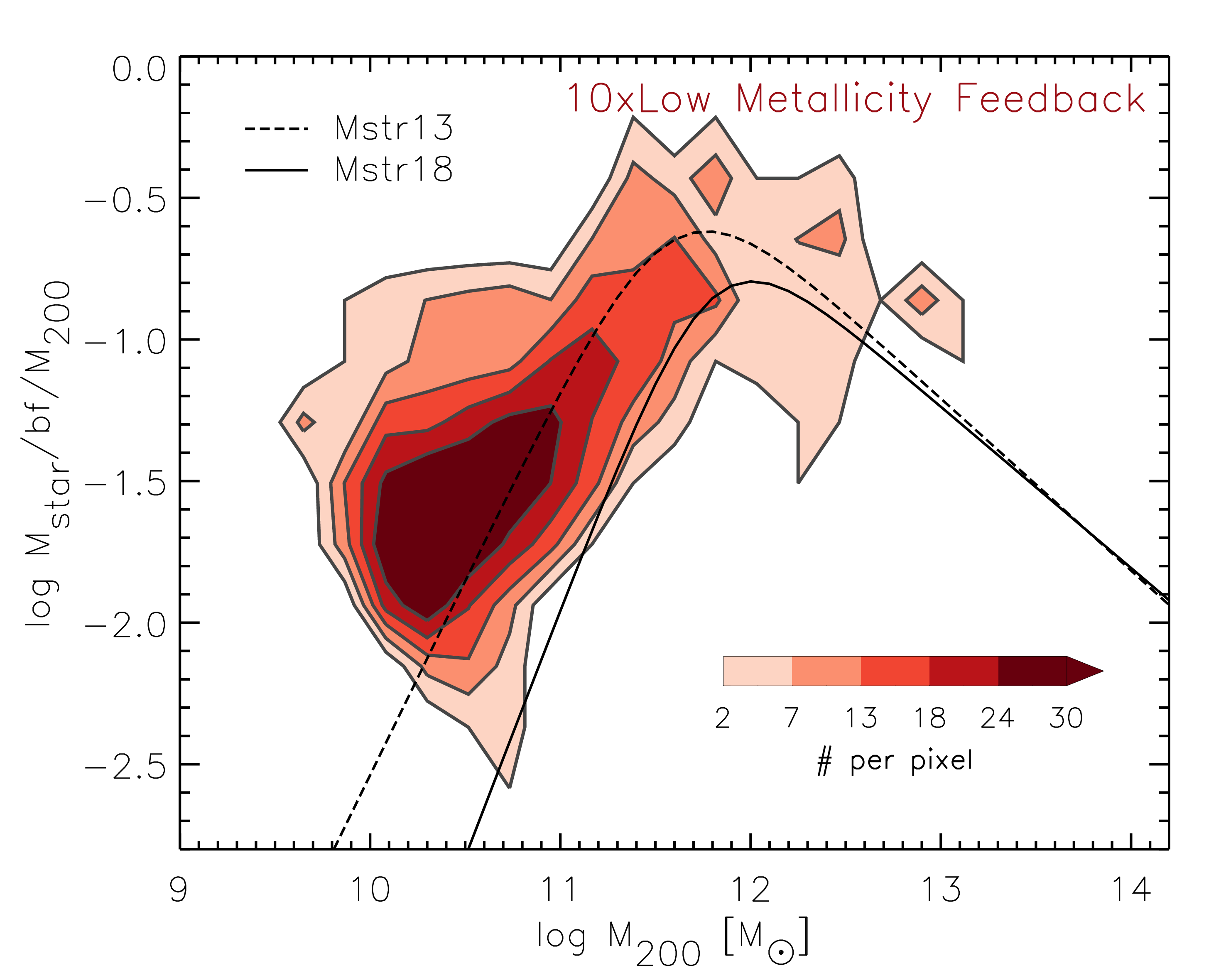}
\includegraphics[width=0.48\textwidth]{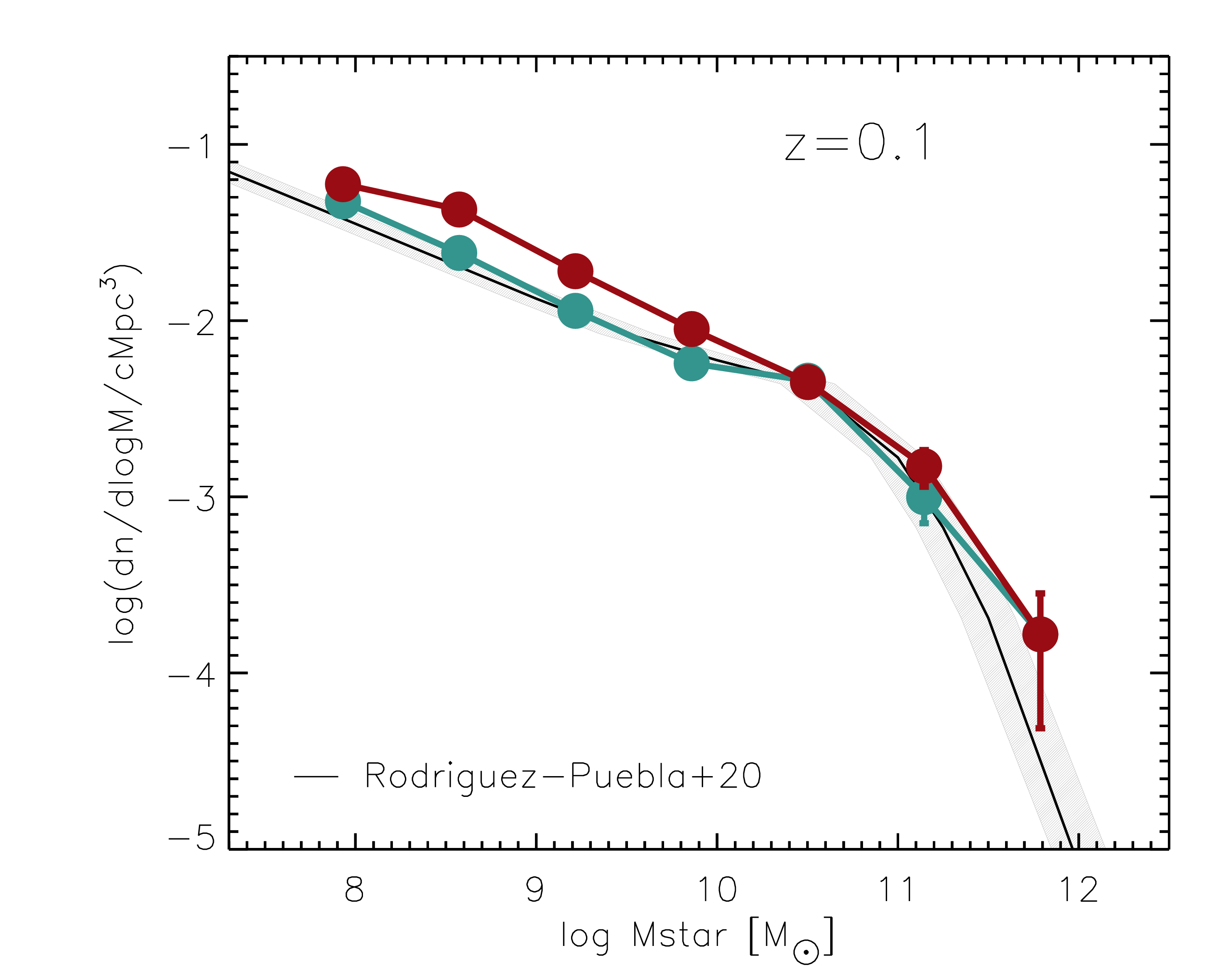}
\caption[width=\textwidth]
{Same as in Fig. \ref{fig:oldsetup1} but comparing the fiducial model with a run where the low metallicity feedback  is reduced by a factor 2.}
\label{fig:oldsetup2}
\end{figure}

Another notable improvement, aligning our model galaxies more closely with observational data, was achieved by increasing the feedback efficiency from low-metallicity stars (LMF). As outlined in \cite{valentini23} \cite[and also in other numerical works as][and references therein]{Maio2010, Schaye2015, Maio2016, Ma2017, Pillepich2018}, this boosted feedback was introduced to simulate the increased feedback energy expected from SNe generated by Population III stars. Specifically, the model star-forming particles with metallicities $\rm{Z<0.05\ Z_\odot}$ produce feedback energy that is boosted with respect to the usual $\rm{E_{SN} =10^{51} erg}$ by a factor determined by a model parameter, which we have doubled in this version of the code and set to 20. Fig. \ref{fig:oldsetup2} illustrates the impact of this choice in one of the simulated boxes. As expected, the SFRD with 10xLMF is higher than in our fiducial (20x) case, particularly at higher redshift. Additionally, it peaks at higher redshifts compared to the case with 20x. The middle panel shows that the excess of SF mainly occurs in galaxies populating small halos, which is also reflected in the excess of low-mass galaxy counts in stellar mass function in the bottom panel.

\section{Estimate of shielding}
\label{app:shield}
\begin{figure}
\centering
\includegraphics[width=\columnwidth]{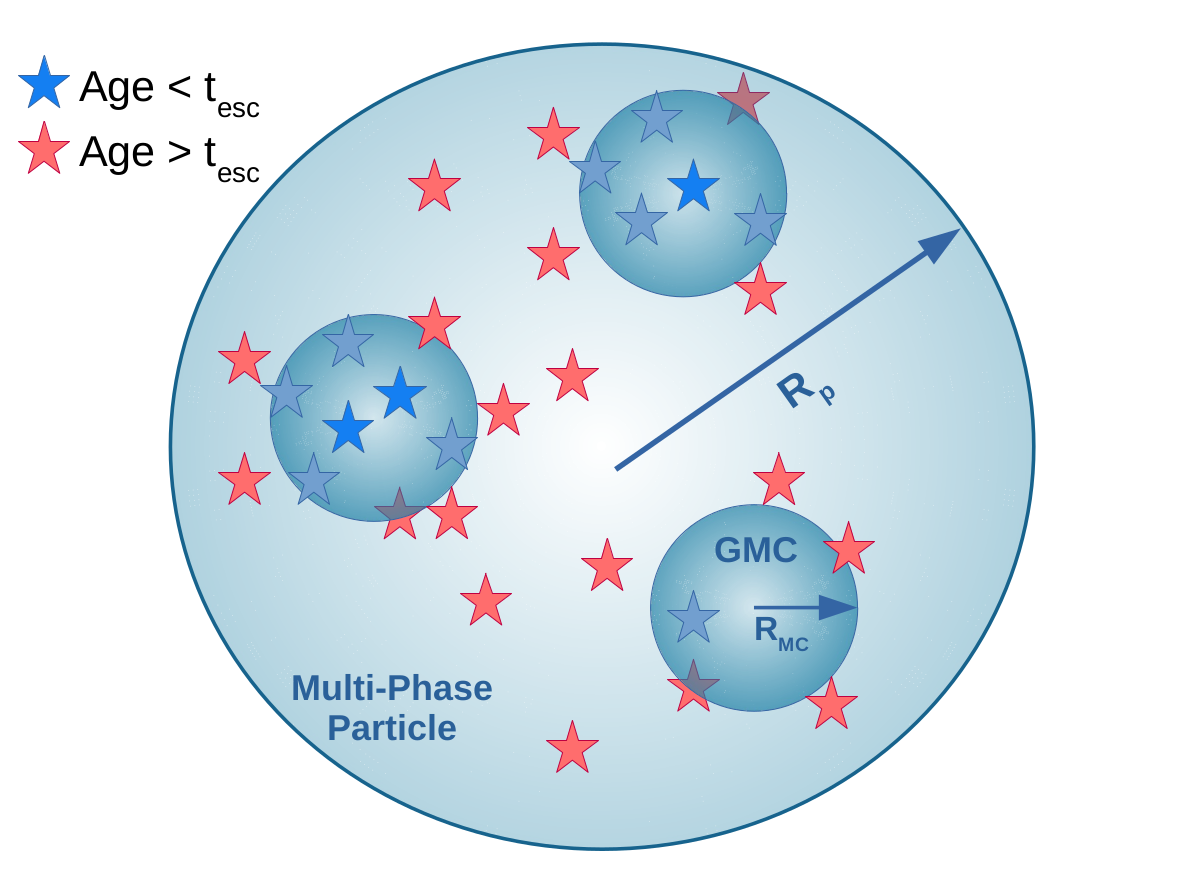}
\caption[width=\textwidth]
{Representation of the idealized molecular mass distribution inside a multi-phase gas particle. The destruction of \hmol is carried out following this schematic. The basic idea is that the \hmol mass (+helium) is organized into Giant Molecular Clouds (GMC), represented by the smaller spheres, with mass $\rm{M_{MC}}$ and radius $\rm{R_{MC}}$. These GMC are considered randomly distributed inside the gas particle, the volume of which is estimated as the ratio between the
particle mass and the SPH density, with a corresponding size $R_p$. Stars younger than a given $\rm{t_{esc}}$ are considered to be still within the GMC, while older stars are considered to be outside.
}
\label{fig:MC}
\end{figure}

To estimate both dust shielding and \hmol self-shielding, we adopt the schematic geometry depicted in Fig. \ref{fig:MC}. The molecular mass is divided into giant molecular clouds (GMC), characterized by a typical mass $M_{MC}$ and radius $R_{MC}$. 
At the resolution adopted in this work, even GMC are about one order of magnitude less massive than individual gas particles.
The GMC are randomly distributed in the particle volume, which, consistently with the MUPPI model, is estimated as the ratio between the particle mass and the SPH density. The corresponding radius is $R_{\rm p}$. 
Moreover, we assume that stars younger than an escape timescale $t_{\rm esc} \sim \mbox{a few Myr}$ (henceforth young stars; in our computations, we set $t_{\rm esc}=3$ Myr  ) are still embedded in the parent GMC whose medium practically completely absorbs their UV radiation; older stars (old stars) are distributed in the particle volume outside the GMC\footnote{Note that this picture is that adopted now by several galaxy SED models including dust reprocessing, following the concept of age-dependent extinction introduced in \cite{silva98}. Adopted values range from $\sim 2$ to $10$ Myr}. 
Therefore, \hmol is affected in a different way by the radiation emitted by stars younger and older than $t_{\rm esc}$. 


The term accounting for destruction in Eq. \ref{eq:newmuppi} is given by the following integral over the volume $V_{aMC}$ of all the $N_{MC}$ molecular clouds:
\begin{equation}
\dot{M}_{\rm{destH_2}}= \int_{V_{aMC}} \rho_{{\rm H}_2}(\Vec{r_g}) \, C_{\rm LW} \, 4\pi\, J_{\nu,{\rm LW}} (\Vec{r_g})\, d^3 r_g,
\label{eq:intdestH2}
\end{equation}
where we use the subscript $g$ (gas)
for the dumb integration position, for reasons that will be clear in the following. We estimate separately the contribution to the LW radiation field $J_{\nu,{\rm LW}}$ arising from the young stars inside the same cloud $J_{\nu ,{\rm LW},\mbox{y} }$ and from old stars outside any cloud $J_{\nu ,{\rm LW},\mbox{o} }$.
The former is given by the following integral over the volume $V_{MC}$ of one of the identical MC:
\begin{align}
\label{eq:intJy}
J_{\nu ,{\rm LW},\mbox{y} }&(\Vec{r_g}) =\\
&\int_{V_{1MC}} 
\frac{L_{\nu ,{\rm LW},\mbox{1y} }}{V_{1MC}} \frac{1}{4\pi \, r_{yg}^2} \exp(-\tau_{d,yg})    S_{H_2}(N_{H_2,yg})
\, d^3 r_y \nonumber.
\end{align}
Here, $L_{\nu ,{\rm LW},\mbox{1y} }$  is the total LW specific luminosity of all young stars inside a given MC. The first fraction represents their constant volume emissivity, estimated assuming a continuous smooth distribution of sources\footnote{We estimate the integrals by discrete Montecarlo sampling. Our results converge for practical purposes when we use more than a few tens of discrete sources.}, $r_{yg} \equiv \Vec{r_y}-\Vec{r_g}$,
$\tau_{d,yg} \equiv \tau_d(|\Vec{r_y}-\Vec{r_g}|)$ is 
the dust optical depth between the positions $\Vec{r_y}$ and $\Vec{r_g}$, and $S_{H_2}(N_{H_2,yg})$ is the \hmol self shielding factor that depends on the \hmol column density $N_{H_2,yg} \equiv N_{H_2}(|\Vec{r_y}-\Vec{r_g}|)$. Using Eq. \ref{eq:intJy} into Eq. \ref{eq:intdestH2}, we obtain the contribution of young stars to \hmol destruction:
\begin{equation}
    \dot{M}_{\rm{destH_2,y}}=
M_{\rm{H}_2}\, C_{LW} \,
L_{\nu ,{\rm LW},\mbox{1y} } 
\times
 \mathcal{A},
\label{eq:destH2y}
\end{equation}
where
\begin{equation}
\mathcal{A} = \frac{1}{V_{1MC}^2}
\int\limits_{V_{1MC}}\int\limits_{V_{1MC}} \frac{\exp(-\tau_{d,yg})     S_{H_2}(N_{H_2,yg})}{r_{yg}^2} 
d^3 r_g d^3 r_y,
\label{eq:calA}
\end{equation}
is the mean over all pairs of points in the MC of the integrand.
To derive the former expression we used $\rho_{{\rm H}_2}(\Vec{r_g}) = M_{\rm{H}_2}/(N_{MC}V_{1MC})$, and the fact  that with our assumptions the integral appearing in Eq. \ref{eq:intdestH2} over the volume of all MC is $N_{MC}$ times the integral over the volume of one MC.

As for the \hmol self shielding factor $S_{H_2}$, we adopt the approximations by \citet{draine_bertoldi1996}:
\begin{equation}
     S_{H_2} (N_{H_2} )= \min \left[1, \,  
     \left(\frac{N_{H_2}}{10^{14}~\rm{cm}^{-2}}\right)^{-0.75} \right].
     \label{eq:selshi}
\end{equation}
We assume the \hmol column density of a GMC to be approximately $N_{H_2} \simeq 1.15 \times 10^{21}$ cm$^{-2}$ from the surface to the center. The column density is proportional to $M_{\rm MC}/R_{\rm MC}^2$. For example, this value is consistent with a GMC having a representative mass of $M_{\rm MC} = 10^6$ M$\odot$ and a radius of $R_{\rm MC} = 100$ pc, assuming a helium fraction of 0.24.
Indeed \cite{miville17} found that half of the MW molecular mass is contained in MC more massive than $8.4 \times 10^5$ M$_\odot$, and a well-defined correlation between the mass and the radius of MC in the MW, yielding a corresponding radius of 96 pc and $N_{H_2}\simeq 1.04  \times 10^{21}$ cm$^{-2}$.
We note, however, that the universality of MC physical properties in galaxies is debated due to observations suggesting a significant increase of  $N_{H_2}$ at higher redshift and SF activity \cite{dessauges19}. Were this the case, our conclusion that LW destruction is practically negligible for the purposes of the present paper would become even stronger.  

The corresponding dust optical depth can be evaluated by recalling that our model of dust evolution approximates the size distribution with just two representative sizes: large grains (L), featuring a radius $a_L \simeq 0.05 \,\mu\mbox{m}$ and small ones (S) with $a_S \simeq 0.005 \,\mu\mbox{m}$. For this section, we can adopt for both silicate and carbonaceous grains an intermediate value of the material density
$s \simeq 3 \mbox{gr cm}^{-2}$ \citep{granato21,parente22}. We approximate the grain cross-section at the LW wavelengths with the geometrical one $\pi a^2$. The column density $N_{d,a}$ of grains of a given size $a$ is related to  $N_{H_2}$ by 
\begin{equation}
    \frac{4}{3} \pi a^3 s \, N_{d,a} = \frac{1}{X} 2 \, m_{H}  N_{H_2} D_a,
\end{equation}
where $D_a$ is the D/G mass ratio, and X is the H mass fraction. Therefore, the corresponding dust optical depth is 
\begin{equation}
\tau_{d,a} \simeq \frac{3 \, m_H N_{H_2} D_a}{2\, s\, X\, a}
\end{equation}
For our two size dust population this yields
\begin{equation}
\tau_{d} \simeq 2.2 \left(\frac{D_L}{0.01}+\frac{10 \, D_S}{0.01} \right) \left(\frac{N_{H_2}}{10^{21} \mbox{cm}^{-2}}\right). 
\label{eq:taud}
\end{equation}
Using Eqs.\ \ref{eq:selshi} and \ref{eq:taud}, and the assumed $N_{H_2}$ along the radius, we can then estimate by a Montecarlo procedure the mean $\mathcal{A}$ (Eq.\ \ref{eq:calA}), as a function of the dust to gas ratio(s) in each MP particle. In practice, we use a table of pre-computed values, well approximated by 
\begin{equation}
    \log (\mathcal{A} \; R_{MC}^2) = -0.03\, x^3-0.12\,x^2-0.17\, x -4.50,
\end{equation}
where $x =\max(\log \tau_d,-2)$.

\hmol can also be affected by the destruction 
from LW radiation coming from outside the MC, generated by stars older than $t_{\rm esc}$ (but as discussed above, younger than $\sim 20$ Myr; older stars do not contribute significantly to the LW region) spread in the whole particle volume. If $L_{\nu ,{\rm LW},\mbox{o}}$ is the total LW luminosity of all old stars in the particle, and $V_P$ is the particle volume outside MC, we can write an expression analogous to Eq. \ref{eq:intJy} to account for the contribution to the radiation field inside a molecular cloud due to all the stars:
\begin{align}
\label{eq:intJyo}
J_{\nu ,{\rm LW},\mbox{o} }&(\Vec{r_g}) =\\
&\int_{V_{P}} 
\frac{L_{\nu ,{\rm LW},\mbox{o} }}{V_{P}} \frac{1}{4\pi \, r_{og}^2} \exp(-\tau_{d,og})    S_{H_2}(N_{H_2,og})
\, d^3 r_o \nonumber,
\end{align}
which inserted into Eq. \ref{eq:intdestH2} yields:
\begin{equation}
\dot{M}_{\rm{destH_2,o}}=
M_{\rm{H}_2}\, C_{LW} \,
L_{\nu ,{\rm LW},\mbox{o} } \,
 \times \mathcal{B},
\label{eq:destH2o}
\end{equation}
where
\begin{equation}
\mathcal{B}  =\frac{1}{V_{aMC}^2V_{P}}
\int_{V_{aMC}}\int_{V_{P}}  
\frac{\exp(-\tau_{d,og})\, S_{H_2}(N_{H_2,og})}{r_{og}^2}      
d^3 r_g d^3 r_o.
\label{eq:calB}
\end{equation}
This expression is similar to Eq. \ref{eq:calA}, but now the mean refers to all pairs of points "og", the first representing an old star outside MC and the second one representing \hmol gas inside the specific MC. To estimate by Montecarlo method this average, we accept the simplifying assumption that the segment connecting the two points does not intersect other MC, and we neglect the contribution to $\tau_{d,og}$ from less dense dusty ISM outside the MC. Again, we notice that giving up to one or both of these simplifications would produce a greater shielding factor, reinforcing our conclusion that LW destruction is negligible for our purposes. 
We found that $\mathcal{B}$ (Eq.\ \ref{eq:calB}) is well approximated by 
\begin{equation}
\log(\mathcal{B} \; R_{MC}^2 ) =-0.28 \, x^2 + (0.10 \, y - 0.69) \, x - 1.71 \,y - 5.24 ,
\end{equation}
where $x =\max(\log \tau_d,-1)$ and $y=\log (R_p/R_{MC})$. This approximation holds for $x<2.5$ and $0.5 < y < 2.5$, ranges more than sufficient to cover the requirements of our simulations.

\end{appendix}
\end{document}